\documentclass[varenna]{cimento}
\usepackage{graphicx,xspace}

\title{Ultracold Fermi gases in the BEC-BCS crossover:\\ a review from the Innsbruck perspective}

\author{Rudolf Grimm}

\institute{Institute of Experimental Physics and Center for Quantum Physics\\
University of Innsbruck, Technikerstra\ss{}e 25, A-6020 Innsbruck, Austria}

\institute{Institute for Quantum Optics and Quantum Information (IQOQI)\\
Austrian Academy of Sciences, Otto-Hittmair-Platz 1, A-6020
Innsbruck, Austria}

\begin{document}

\maketitle


\newcommand{\Li}{$^6$Li\xspace}
\newcommand{\K}{$^{40}$K\xspace}
\newcommand{\Eb}{$E_{\rm b}$\xspace}
\newcommand{\kB}{$k_{\rm B}$\xspace}
\newcommand{\EF}{$E_{\rm F}$\xspace}
\newcommand{\kF}{$k_{\rm F}$\xspace}
\newcommand{\TF}{$T_{\rm F}$\xspace}
\newcommand{\kBT}{$k_{\rm B}T$\xspace}
\newcommand{\kFa}{$1/k_{\rm F}a$\xspace}

\section{Introduction}

By the time of the ``Enrico Fermi'' Summer School in June 2006, quantum
degeneracy in ultracold Fermi gases has been reported by 13 groups worldwide
\cite{Demarco1999oof, Truscott2001oof, Schreck2001qbe, Granade2002aop,
Hadzibabic2002tsm, Roati2002fbq, Jochim2003bec, Kohl2005fai, Silber2005qdm,
Ospelkaus2006idd, Aubin2006rsc, Mcnamara2006dbf, yb173}. The field is rapidly
expanding similar to the situation of Bose-Einstein condensation at the time of
the ``Enrico Fermi'' Summer School in 1998 \cite{Varenna1998}. The main two
species for the creation of ultracold Fermi gases are
the alkali atoms potassium ($^{40}$K) \cite{Demarco1999oof, Roati2002fbq,
Kohl2005fai, Ospelkaus2006idd, Aubin2006rsc} and lithium ($^6$Li)
\cite{Truscott2001oof, Schreck2001qbe, Granade2002aop, Hadzibabic2002tsm,
Jochim2003bec, Silber2005qdm}. At the time of the School, degeneracy was
reported for two new species, $^3$He$^*$ \cite{Mcnamara2006dbf} and $^{173}$Yb
\cite{yb173}, adding metastable and rare earth species to the list.

Fermionic particles represent the basic building blocks of matter,
which connects the physics of interacting fermions to very
fundamental questions. Fermions can pair up to form composite
bosons. Therefore, the physics of bosons can be regarded as a
special case of fermion physics, where pairs are tightly bound and
the fermionic character of the constituents is no longer relevant.
This simple argument already shows that the physics of fermions is
in general much richer than the physics of bosons.

Systems of interacting fermions are found in many areas of
physics, like in condensed-matter physics (e.g.\ superconductors),
in atomic nuclei (protons and neutrons), in primordial matter
(quark-gluon plasma), and in astrophysics (white dwarfs and
neutron stars). Strongly interacting fermions pose great
challenges for many-body quantum theories. With the advent of
ultracold Fermi gases with tunable interactions and controllable
confinement, unique model systems have now become experimentally
available to study the rich physics of fermions.

In this contribution, we will review a series of experiments on
ultracold, strongly interacting Fermi gases of $^6$Li which we
conducted at the University of Innsbruck. We will put our
experiments into context with related work and discuss them
according to the present state-of-the art knowledge in the field.
After giving a brief overview of experiments on strongly
interacting Fermi gases (Sec.~\ref{sec_overview}), we will discuss
the basic interaction properties of $^6$Li near a Feshbach
resonance (Sec.~\ref{sec_feshbach}). Then we will discuss the main
experimental results on the formation and Bose-Einstein
condensation of weakly bound molecules (Sec.~\ref{sec_mBEC}), the
crossover from a molecular Bose-Einstein condensate 
to a fermionic superfluid (Sec.~\ref{sec_crossover}), and detailed
studies on the crossover by collective modes
(Sec.~\ref{sec_oscillations}) and pairing-gap spectroscopy
(Sec.~\ref{sec_gap}).


\section{Brief history of experiments on strongly interacting Fermi
gases} \label{sec_overview}


To set the stage for a more detailed presentation of our results,
let us start with a brief general overview of the main
experimental developments in the field of ultracold, strongly
interacting Fermi gases; see also the contributions by D.\ Jin and
W. Ketterle in these proceedings. The strongly interacting regime
is realized when the scattering length, characterizing the
two-body interacting strength, is tuned to large values by means
of Feshbach resonances \cite{Tiesinga1993tar,Inouye1998oof}. In
the case of Bose gases with large scattering lengths rapid
three-body decay \cite{Roberts2000mfd,Weber2003tbr,Kraemer2006efe}
prevents the experiments to reach the strongly interacting
regime~\footnote{This statement refers to macroscopically trapped
gases of a large number of atoms. Highly correlated systems of
bosons can be created in optical lattices \cite{Greiner2002qpt}.}.
Experiments with ultracold Fermi gases thus opened up a door to
the new, exciting regime of many-body physics with ultracold
gases.

The creation of a strongly interacting Fermi gas was first
reported in 2002 by the group at Duke University
\cite{Ohara2002ooa}. They studied the expansion of a \Li gas with
resonant interactions after release from the trap and observed
hydrodynamic behavior. In similar experiments, the group at the
ENS Paris provided measurements of the interaction energy of
ultracold \Li in the strongly interacting region
\cite{Bourdel2003mot}.

In 2003 ultracold diatomic molecules entered the stage. Their
formation is of particular importance in atomic Fermi gases, as
their bosonic nature is connected with a fundamental change of the
quantum statistics of the gas. The JILA group demonstrated
molecule formation in an ultracold Fermi gas of \K
\cite{Regal2003cum}, followed by three groups working with \Li:
Rice University \cite{Strecker2003coa}, the ENS Paris
\cite{Cubizolles2003pol}, and Innsbruck University
\cite{Jochim2003pgo}. The latter experiments on \Li also
demonstrated an amazing fact. Molecules made of fermionic atoms
can be remarkably stable against inelastic decay, allowing for the
formation of stable molecular quantum gases.

In late 2003 three groups reported on the achievement of molecular
Bose-Einstein condensation, our group (\Li) \cite{Jochim2003bec},
the JILA group (\K) \cite{Greiner2003eoa}, and the MIT group (\Li)
\cite{Zwierlein2003oob}, followed early in 2004 by the ENS group
(\Li) \cite{Bourdel2004eso}.
Early in 2004, the JILA group \cite{Regal2004oor} and the MIT
group \cite{Zwierlein2004cop} demonstrated pair condensation in
strongly interacting Fermi gases with resonant interactions, i.e.\
beyond the BEC regime. These experiments demonstrated a new
macroscopic quantum state of ultracold matter beyond
well-established BEC physics, which has stimulated an enormous
interest in the field.

The experiments then started to explore the crossover from a
BEC-type system to a fermionic superfluid with
Bardeen-Cooper-Schrieffer (BCS) type pairing.  Elementary
properties of the Fermi gas in the BEC-BCS crossover were studied
by several groups. In Innsbruck, we showed that the crossover
proceeds smoothly and can be experimentally realized in an
adiabatic and reversible way \cite{Bartenstein2004cfa}. At the ENS
the crossover was investigated in the free expansion of the gas
after release from the trap. Measurements of collective excitation
modes at Duke University and in Innsbruck showed exciting
observations and provided pieces of evidence for superfluidity in
the strongly interacting gas. The Duke group measured very low
damping rates, which could not be explained without invoking
superfluidity \cite{Kinast2004efs}. Our work on collective
oscillations \cite{Bartenstein2004ceo} showed a striking breakdown
of the hydrodynamic behavior of the gas when the interaction
strength was changed, suggesting a superfluid-normal transition.

Spectroscopy on fermionic pairing based on a radio-frequency
method showed the ``pairing gap'' of the strongly interacting gas
along the BEC-BCS crossover \cite{Chin2004oop}. In these
experiments performed in Innsbruck, temp\-era\-ture-dependent
spectra suggested that the resonantly interacting Fermi gas was
cooled down deep into the superfluid regime. A molecular probe
technique of pairing developed at Rice University provided clear
evidence for pairing extending through the whole crossover into
the weakly interacting BCS regime \cite{Partridge2005mpo}.
Measurements of the heat capacity of the strongly interacting gas
performed at Duke University showed a transition at a temperature
where superfluidity was expected \cite{Kinast2005hco}. After
several pieces of experimental evidence provided by different
groups, the final proof of superfluidity in strongly interacting
Fermi gases was given by the MIT group in 2005
\cite{Zwierlein2005vas}. They observed vortices and vortex arrays
in a strongly interacting Fermi gas in various interaction
regimes.

New phenomena were recently explored in studies on imbalanced
spin-mixtures at Rice University \cite{Partridge2005pap} and at
MIT \cite{Zwierlein2005fsw}. These experiments have approached a
new frontier, as such systems may offer access to novel superfluid
phases. Experiments with imbalanced spin-mixtures also revealed
the superfluid phase transition in spatial profiles of the
ultracold cloud \cite{Zwierlein2006doo}.

\section{Interactions in a $^6$Li spin mixture}
\label{sec_feshbach}

Controllable interactions play a crucial role in all experiments
on strongly interacting Fermi gases. To exploit an $s$-wave
interaction at ultralow temperatures, non-identical particles are
needed; thus the experiments are performed on mixtures of two
different spin states. Feshbach resonances
\cite{Tiesinga1993tar,Inouye1998oof,Loftus2002rco} allow tuning
the interactions through variations of an external magnetic field.
In this section, we review the two-body interaction properties of
$^6$Li. In particular, we discuss the behavior close to a wide
Feshbach resonance with very favorable properties for interaction
tuning in strongly interacting Fermi gases.

\subsection{Energy levels of $^6$Li atoms in a magnetic field}
\label{ssec_paschenback}

\begin{figure}
\begin{center}
\includegraphics[angle=0,width=10cm]{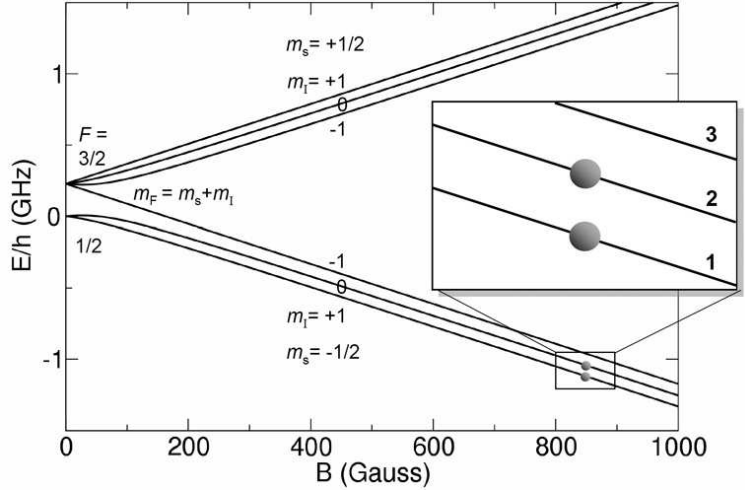}
\end{center}
\caption{Energy levels for the electronic ground state of $^6$Li
atoms in a magnetic field. The experiments on strongly interacting
Fermi gases are performed in the high magnetic field range, where
the nuclear spin essentially decouples from the electron spin. The
two-component atomic mixture is created in the lowest two states,
labelled with 1 and 2 (inset), close to the broad Feshbach
resonance centered at 834\,G.} \label{paschenback}
\end{figure}

The magnetic-field dependence of the energy structure of $^6$Li
atoms in the electronic S$_{1/2}$ ground state is shown in
Fig.~\ref{paschenback}. The general behavior is similar to any
alkali atom \cite{Arimondo1977edo} and is described by the
well-known Breit-Rabi formula. At zero magnetic field, the
coupling of the $^6$Li nuclear spin ($I=1$) to the angular
momentum of the electron ($J=1/2$) leads to the hyperfine
splitting of 228.2\,MHz between the states with quantum numbers
$F=I + J$ and $F=I - J$.

Already at quite moderate magnetic fields the Zeeman effect turns
over into the high-field regime, where the Zeeman energy becomes
larger than the energy of the hyperfine interaction. Here the
nuclear spin essentially decouples from the electron spin. In
atomic physics this effect is well known as the ``Paschen-Back
effect of the hyperfine structure'' or ``Back-Goudsmit effect''
\cite{Arimondo1977edo}. In the high-field region the states form
two triplets, depending on the orientation of the electron spin
($m_s = \pm 1/2$), where the states are characterized by the
orientation of the nuclear spin with quantum number $m_I$. For
simplicity, we label the states with numbers according to
increasing energy (see inset in Fig.~\ref{paschenback}). The
lowest two states 1 and 2 are of particular interest for creating
stable spin mixtures. These two states $m_s=-1/2, m_I=+1$
($m_s=-1/2, m_I=0$) are adiabatically connected with the states
$F=1/2, m_F=1/2$ ($F=1/2, m_F=-1/2$) at low magnetic fields.

\subsection{Tunability at the marvelous 834\,G Feshbach resonance}
\label{ssec_tunability}

\begin{figure}
\vspace{3mm}
\begin{center}
\includegraphics[width=8cm]{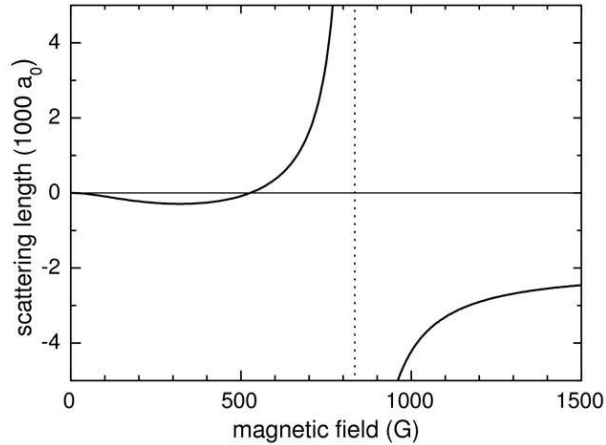}
\vspace{-3mm}
\end{center}
\caption{Tunability of the $s$-wave interactions in a spin mixture
of $^6$Li atoms in the two lowest spin states 1 and 2. The
$s$-wave scattering length $a$ shows a pronounced, broad resonance
as a function of the magnetic field
\cite{Houbiers1998eai,Bartenstein2005pdo}. The vertical dotted
line indicates the exact resonance field (834\,G) where $a$ goes
to infinity and the interaction is only limited through
unitarity.} \label{resonance}
\end{figure}

Interactions between $^6$Li atoms in states 1 and 2 show a
pronounced resonance in $s$-wave scattering
\cite{Houbiers1998eai,Bartenstein2005pdo} with favorable
properties for the experiments on strongly interacting Fermi
gases. Fig.~\ref{resonance} displays the scattering length $a$ as
a function of the magnetic field $B$. The center of the resonance,
i.e.\ the point where $a$ diverges, is located at 834\,G. This
resonance center is of great importance to realize the
particularly interesting situation of a universal Fermi gas in the
unitarity limit; see discussion in \ref{ssec_universal}.


The investigation of the broad Feshbach resonance in $^6$Li has a
history of almost ten years. In 1997, photoassociation
spectroscopy performed at Rice University revealed a triplet
scattering length that is negative and very large
\cite{Abraham1997tsw}. A theoretical collaboration between Rice
and the Univ.\ of Utrecht \cite{Houbiers1998eai} then led to the
prediction of the resonance near 800\,G. In 2002, first
experimental evidence for the resonance was found at MIT
\cite{Dieckmann2002doa}, at Duke University \cite{Ohara2002mot},
and in Innsbruck \cite{Jochim2002mfc}. At about 530\,G,
experiments at Duke and in Innsbruck showed the zero crossing of
the scattering length that is associated with the broad resonance.
The MIT group observed an inelastic decay feature in a broad
magnetic-field region around 680\,G. The decay feature was also
observed at the ENS Paris, but at higher fields around 720\,G
\cite{Bourdel2003mot}. The ENS group also reported indications of
the resonance position being close to 800\,G. In Innsbruck the
decay feature was found \cite{JochimPhD} in a broad region around
640\,G~\footnote{The interpretation of these inelastic decay
features involves different processes, which depend on the
particular experimental conditions, see also
\ref{ssec_equilibrium}. In a three-body recombination event,
immediate loss occurs when the release of molecular binding energy
ejects the particles out of the trap. Another mechanism of loss is
vibrational quenching of trapped, weakly bound molecules. The fact
that, in contrast to bosonic quantum gases, maximum inelastic
decay loss does not occur at the resonance point, but somewhere in
the region of positive scattering length is crucial for the
stability of strongly interacting Fermi gases with resonant
interactions.}. Molecule dissociation experiments at MIT
\cite{Zwierlein2004cop,Schunck2005fri} provided a lower bound of
822\,G for the resonance point. To date the most accurate
knowledge on $a(B)$ in $^6$Li spin mixtures results from an
experiment-theory collaboration between Innsbruck and NIST on
radio-frequency spectroscopy on weakly bound molecules
\cite{Bartenstein2005pdo}. This work puts the resonance point to
834.1\,G within an uncertainty of $\pm1.5$\,G.

The dependence $a(B)$ near the Feshbach resonance can be
conveniently described by a fit formula \cite{Bartenstein2005pdo},
which approximates the scattering length  in a range between 600
and 1200\,G to better than 99\%,
\begin{equation}
a(B) = a_{\rm bg} \left( 1 + \frac{\Delta B}{B-B_0} \right)
\left(1+\alpha(B-B_0)\right)
\end{equation}
with $a_{\rm bg}= -1405\,a_0$, $B_0 = 834.15$\,G, $\Delta B
=300$\,G, and $\alpha = 0.040$\,kG$^{-1}$; here $a_0 =
0.529177$\,nm is Bohr's radius.

Concerning further Feshbach resonances in $^6$Li, we note that
besides the broad 834\,G resonance in the $(1,2)$ spin mixture,
similar broad $s$-wave resonances are found in $(1,3)$ and in
$(2,3)$ mixtures with resonance centers at 690\,G and at 811\,G,
respectively \cite{Bartenstein2005pdo}. The $(1,2)$ spin mixture
also features a narrow Feshbach resonance near 543\,G with a width
of roughly 100\,mG \cite{Strecker2003coa,Schunck2005fri}.
Moreover, Feshbach resonances in $p$-wave scattering of $^6$Li
have been observed in $(1,1)$, $(1,2)$, and $(2,2)$ collisions at
the ENS \cite{Zhang2004pwf} and at MIT \cite{Schunck2005fri}.

\subsection{Weakly bound dimers}
\label{ssec_dimers}

A regime of particular interest is realized when the scattering
length $a$ is very large and positive. The scale for ``very
large'' is set by the van der Waals interaction between two $^6$Li
atoms, characterized by a length $R_{\rm vdW} = (m
C_6/\hbar^2)^{1/4}/2 = 31.26$\,$a_0$ (for \Li, $C_6=1393$\,a.u.\
and the atomic mass is $m = 6.015\,$u). For $a \gg R_{\rm vdW}$, a
weakly bound molecular state exists with a binding energy given by
the universal formula~\footnote{A useful correction to the
universal expression for the non-zero range of the van der Waals
potential is $E_{\rm b} = \hbar^2/(m (a-\bar{a})^2)$ where
$\bar{a}=0.956\,R_{\rm vdW}$ \cite{Gribakin1993cot}.}
\begin{equation}
E_{\rm b} = \frac{\hbar^2}{m a^2}. \label{eq_eb}
\end{equation}
In this regime, the molecular wave function extends over a much
larger range than the interaction potential and, for large
interatomic distances $r \gg R_{\rm vdW}$, falls off exponentially
as $\exp(-r/a)$. The regime, in which a bound quantum object is
much larger than a classical system, is also referred to as the
``quantum halo regime'' \cite{Jensen2004sar}. For quantum halo
states, the details of the short-range interaction are no longer
relevant and the physics acquires universal character
\cite{Braaten2006uif}. Here two-body interactions are completely
characterized by $a$ as a single parameter.

\begin{figure}
\begin{center}
\includegraphics[width=8cm]{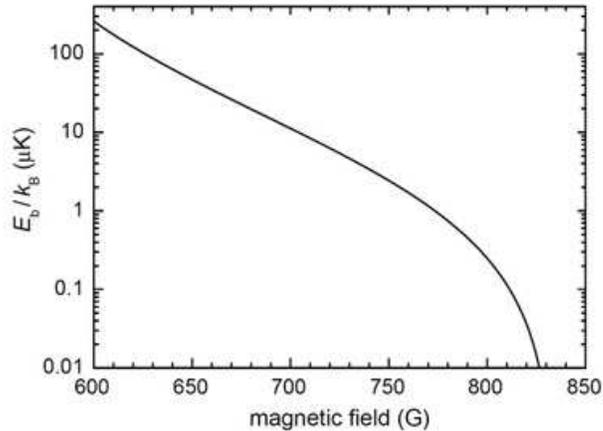}
\vspace{-3mm}
\end{center}
\caption{Binding energy $E_{\rm b}$ of weakly bound $^6$Li
molecules, which exist on the lower side of the 834\,G Feshbach
resonance. Here we use temperature units ($k_{\rm B} \times
1\,\mu{\rm K} \approx h \times 20.8$\,kHz) for a convenient
comparison with our experimental conditions.}
\label{bindingenergy}
\end{figure}

From these considerations, we understand that the lower side of
the 834\,G Feshbach resonance in $^6$Li is associated with the
regime of weakly bound (quantum halo) molecules. The binding
energy of the weakly bound $^6$Li molecular state is plotted in
Fig.~\ref{bindingenergy} as a function of the magnetic field.

Weakly bound molecules made of {\it fermionic} atoms exhibt
striking scattering properties \cite{Petrov2004wbm}. As a big
surprise, which enormously boosted the field of ultracold fermions
in 2003, these dimers turned out to be highly stable against
inelastic decay in atom-dimer and dimer-dimer collisions. The
reason for this stunning behavior is a Pauli suppression effect.
The collisional quenching of a weakly bound dimer to a lower bound
state requires a close encounter of three particles. As this
necessarily involves a pair of identical fermions the process is
Pauli blocked. The resulting collisional stability is in sharp
contrast to weakly bound dimers made of bosonic atoms
\cite{Donley2002amc,Herbig2003poa,Durr2004oom,Xu2003foq}, which
are very sensitive to inelastic decay. The amazing properties of
weakly bound dimers made of fermions were first described in
Ref.~\cite{Petrov2004wbm}. Here we just summarize the main
findings, referring the reader to the lecture of G.~Shlyapnikov in
these proceedings for more details.

For elastic atom-dimer and dimer-dimer collisions, Petrov et al.\
\cite{Petrov2004wbm} calculated the scattering lengths
\begin{eqnarray}
a_{\rm ad} & = & 1.2\,a, \\
a_{\rm dd} & = & 0.6\,a, \label{eq_add}
\end{eqnarray}
respectively. Inelastic processes, described by the loss-rate
coefficients $\alpha_{\rm ad}$ and $\alpha_{\rm dd}$, follow the
general scaling behavior \cite{Petrov2004wbm}
\begin{eqnarray}
\alpha_{\rm ad} & = & c_{\rm ad} \frac{\hbar R_{\rm vdW}}{m}
\left(\frac{R_{\rm vdW}}{a}\right)^{3.33}, \\
\alpha_{\rm dd} & = & c_{\rm dd} \frac{\hbar R_{\rm vdW}}{m}
\left(\frac{R_{\rm vdW}}{a}\right)^{2.55}.
\end{eqnarray}
Here the dimensionless coefficients $c_{\rm ad}$ and $c_{\rm dd}$
depend on non-universal short-range physics. We point out that,
for typical experimental conditions in molecular BEC experiments
(see Sec.~\ref{sec_mBEC}), the factor $(R_{\rm vdW}/a)^{2.55}$
results in a gigantic suppression of five orders of magnitude in
inelastic dimer-dimer collisions!

The general scaling behavior of inelastic loss is universal and
should be the same for \Li and \K, consistent with measurements on
both species \cite{Cubizolles2003pol,Jochim2003pgo,Regal2004lom}
The pre-factors $C_{\rm ad}$ and $C_{\rm dd}$, however, are
non-universal as they depend on short-range three-body physics. A
comparison of the experiments on both species shows that inelastic
decay of weakly bound molecules is typically two orders of
magnitude faster for \K than for \Li. This difference can be
attributed to the larger van der Waals length of $^{40}$K in
combination with its less favorable short-range interactions.

This important difference in inelastic decay is the main reason
why experiments on \Li and \K follow different strategies for the
creation of degenerate Fermi gases. In \Li, the regime of weakly
bound dimers on the molecular side of the Feshbach resonance opens
up a unique route into deep degeneracy, as we will discuss in the
following Section.

\section{The molecular route into Fermi degeneracy: creation of a molecular Bose-Einstein condensate}
\label{sec_mBEC}


In experiments on $^6$Li gases, a molecular Bose-Einstein
condensate (mBEC) can serve as an excellent starting point for the
creation of strongly interacting Fermi gases in the BEC-BCS
crossover regime. In this section, after discussing the various
approaches followed by different groups, we describe the strategy
that we follow in Innsbruck to create the mBEC.

\subsection{A brief review of different approaches}

The experiments on strongly interacting gases of $^6$Li in
different laboratories (in alphabetical order: Duke University,
ENS Paris, Innsbruck University, MIT, Rice University) are based
on somewhat different approaches. The first and the final stages
of all experiments are essentially the same. In the first stage,
standard laser cooling techniques \cite{Metcalf1999book} are
applied to decelerate the atoms in an atomic beam and to
accumulate them in a magneto-optic trap (MOT); for a description
of our particular setup see
Refs.~\cite{Schunemann1998mot,JochimPhD}. In the final stage,
far-detuned optical dipole traps \cite{Grimm2000odt} are used to
store and manipulate the strongly interacting spin mixture. The
creation of such a mixture requires trapping in the high-field
seeking spin states 1 and 2 (see Fig.~\ref{paschenback}), which
cannot be achieved magnetically. The main differences in the
experimental approaches pursued in the five laboratories concern
the intermediate stages of trapping and cooling. The general
problem is to achieve an efficient loading of many $^6$Li atoms
into the small volume of a far-detuned optical dipole trap.

At Rice Univ., ENS, and MIT, magnetic traps are used as an
intermediate stage
\cite{Truscott2001oof,Schreck2001qbe,Hadzibabic2002tsm}. This
approach offers the advantage of a large volume and efficient
transfer from a MOT  with minimum loading losses. To achieve
efficient cooling in the magnetic trap, the experiments then use
bosonic atoms as a cooling agent. At Rice Univ.\ and at ENS, the
$^6$Li atoms are trapped together with the bosonic isotope $^7$Li
\cite{Truscott2001oof,Schreck2001qbe}. The isotope mixture can be
efficiently cooled to degeneracy by radio-frequency induced
evaporation. Finally the sample is loaded into an optical dipole
trap, and the atoms are transferred from their magnetically
trappable, low-field seeking spin state into the high-field
seeking states 1 and 2. The internal transfer is achieved through
microwave and radio-frequency transitions. In this process it is
important to create an incoherent spin mixture, which requires
deliberate decoherence in the sample. At MIT the approach is
basically similar \cite{Hadzibabic2002tsm}, but a huge BEC of Na
atoms is used as the cooling agent. This results in an
exceptionally large number of atoms in the degenerate Fermi gas
\cite{Hadzibabic2003fii}. In all three groups (Rice, ENS, MIT),
final evaporative cooling is performed on the strongly interacting
spin mixture by reducing the power of the optical trap.

The experiments at Duke University \cite{Ohara2002ooa} and in
Innsbruck \cite{Jochim2003bec} proceed in an all-optical way
without any intermediate magnetic traps. To facilitate direct
loading from the MOT, the optical dipole traps used in these
experiments have to start with initially very high laser power.
For the final stage of the evaporation much weaker traps are
needed. Therefore, the all-optical approach in general requires a
large dynamical range in the optical trapping power. The Duke
group uses a powerful 100-W CO$_2$ laser source
\cite{Granade2002aop} both for evaporative cooling and for the
final experiments. In Innsbruck we employ two different optical
trapping stages to optimize the different phases of the
experiment.

\subsection{The all-optical Innsbruck approach}

\begin{figure}
\begin{center}
\includegraphics{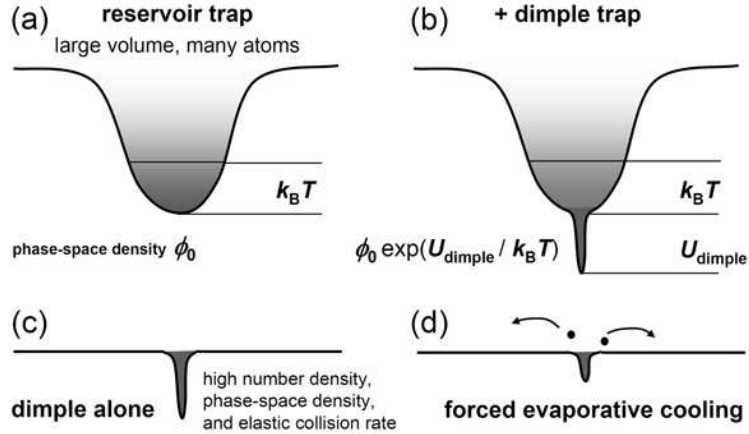}
\end{center}
\caption{Illustration of the ``dimple trick''. The atoms are first
transferred from the MOT into a large-volume optical reservoir
trap (a), here implemented inside of an optical resonator. A
narrow ``dimple'' potential (b) is then added and thermalization
leads to a huge increase of the local density and phase-space
density according to the Boltzmann factor as the temperature is
set by the reservoir. After removal of the reservoir trap (c) one
obtains a very dense sample optically trapped sample. Forced
evaporative cooling can then be implemented (d) by ramping down
the trap power. The dimple trick, originating from work in
Refs.~\cite{Pinkse1997act} and \cite{Stamperkurn1998rfo}, has
proven a very powerful tool for the all-optical creation of
degenerate quantum gases \cite{Weber2003bec,Rychtarik2004tdb}.}
\label{dimpletrick}
\end{figure}

An efficient transfer of magneto-optically trapped lithium atoms
into an optical dipole trap is generally much more difficult than
for the heavy alkali atoms. The much higher temperatures of
lithium in a MOT of typically a few hundred microkelvin
\cite{Schunemann1998mot} require deep traps with a potential depth
of the order of 1\,mK. We overcome this bottleneck of dipole trap
loading by means of a deep large-volume dipole trap serving as a
``funnel''. The trap is realized inside a build-up cavity
constructed around the glass cell \cite{Mosk2001red}. The linear
resonator enhances the power of a 2-W infrared laser (Nd:YAG at a
wavelength of 1064\,nm) by a factor of $\sim$150 and, with a
Gaussian beam waist of 160$\mu$m, allows us to create a 1\,mK deep
optical standing-wave trapping potential. Almost 10$^7$ atoms in
the lower hyperfine level with $F=1/2$ can be loaded from the \Li
MOT  into the resonator-enhanced dipole trap at a temperature of
typically a few 100\,$\mu$K. Note that, when loaded from the MOT,
the spin mixture of states 1 and 2 in the optical dipole trap is
incoherent from the very beginning.

Then we apply a single beam from a 10-W near-infrared laser
(wavelength 1030\,nm), which is focussed to a waist of typically a
few ten $\mu$m ($\sim$25\,$\mu$m in our earlier experiments
\cite{Jochim2003bec,Bartenstein2004cfa,Bartenstein2004ceo},
$\sim$50$\mu$m in more recent work \cite{Altmeyer2006pmo}),
overlapping it with the atom cloud in the standing-wave trapping
potential. The total optical potential can then be regarded as a
combination of a large-volume ``reservoir'' trap in combination
with a narrow ``dimple'' potential. The dimple is efficiently
filled through elastic collisions resulting in a large increase in
local density, phase-space density, and elastic collisions rate;
this ``dimple trick'' is illustrated in Fig.~\ref{dimpletrick}.
After removal of the reservoir, i.e.\ turning off the
standing-wave trap, we obtain a very dense cloud of $\sim$$1.5
\times 10^6$ atoms at a temperature $T \approx 80\,\mu$K, a peak
density of $\sim$$10^{14}$\,cm$^{-1}$, a peak phase-space density
of $5 \times 10^{-3}$, 
and a very
high elastic collision rate of $5 \times 10^{4}$\,s$^{-1}$. In
this way, excellent starting conditions are realized for
evaporative cooling.

A highly efficient evaporation process is then forced by ramping
down the laser power by typically three orders of magnitude within
a few seconds. The formation of weakly bound molecules turns out
to play a very favorable role in this process and eventually leads
to the formation of a molecular BEC. The details of this amazing
process will be elucidated in the following.

%

\subsection{Formation of weakly bound molecules}
\label{ssec_equilibrium}

\begin{figure}
\begin{center}
\includegraphics{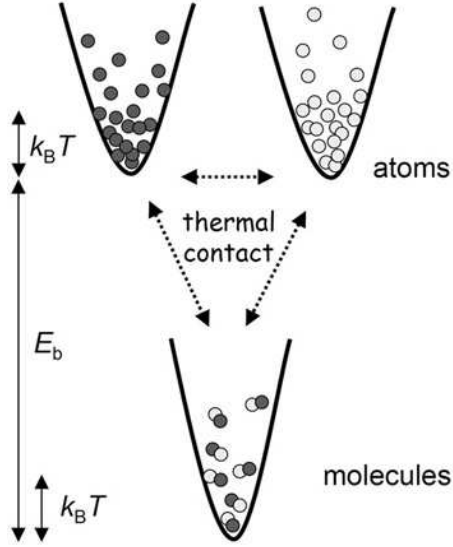}
\end{center}
\caption{Illustration of the atom-molecule thermal equilibrium in
a trapped \Li gas at the molecular side of the Feshbach resonance.
Atoms in the two spin states and molecules represent three
sub-ensembles in thermal contact (thermal energy \kBT). The
molecules are energetically favored because of the binding energy
\Eb, which is reflected in the Boltzmann factor in
Eq.~\ref{eq_equilibrium}. The equilibrium can also be understood
in terms of a balance of the chemical processes of exoergic
recombination and endoergic dissociation \cite{Chin2004tea}.}
\label{equilibrium}
\end{figure}

The formation of weakly bound molecules in a {\it chemical
atom-molecule equilibrium} \cite{Chin2004tea,Kokkelmans2004dam}
plays an essential role in the evaporative cooling process; see
illustration in Fig.~\ref{equilibrium}. In the \Li gas, molecules
are formed through three-body recombination. As the molecular
binding energy \Eb is released into kinetic energy, this process
is exoergic and thus leads to heating of the sample~\footnote{The
relation of released binding energy \Eb to the trap depth is
crucial whether the recombination products remain trapped and
further participate in the thermalization processes. For low trap
depth the recombination leads to immediate loss. This explains why
the loss features observed by different groups
\cite{Dieckmann2002doa,Bourdel2003mot,JochimPhD} shift towards
lower fields at higher trap depths.}. The inverse chemical process
is dissociation of molecules through atom-dimer and dimer-dimer
collisions. These two-body processes are endoergic and can only
happen when the kinetic energy of the collision partners is
sufficient to break up the molecular bond. From a balance of
recombination (exoergic three-body process) and dissociation
(endoergic two-body processes) one can intuitively understand that
molecule formation is favored at low temperatures and high number
densities, i.e.\ at high phase-space densities.

For a non-degenerate gas, the atom-molecule equilibrium follows a
simple relation \cite{Chin2004tea}
\begin{equation}
\phi_{\rm mol} = \phi_{\rm at}^2 \exp\left(\frac{E_{\rm b}}{k_{\rm
B}T}\right), \label{eq_equilibrium}
\end{equation}
where $\phi_{\rm mol}$ and $\phi_{\rm at}$ denote the molecular
and atomic phase-space densities, respectively. The Boltzmann
factor enhances the fraction of molecules in a trapped sample and
can (partially) compensate for a low atomic phase-space density.
Including the effect of Fermi degeneracy, the thermal
atom-molecule equilibrium was theoretically investigated in
Ref.~\cite{Kokkelmans2004dam}.

\begin{figure}
\begin{center}
\includegraphics[width=8cm]{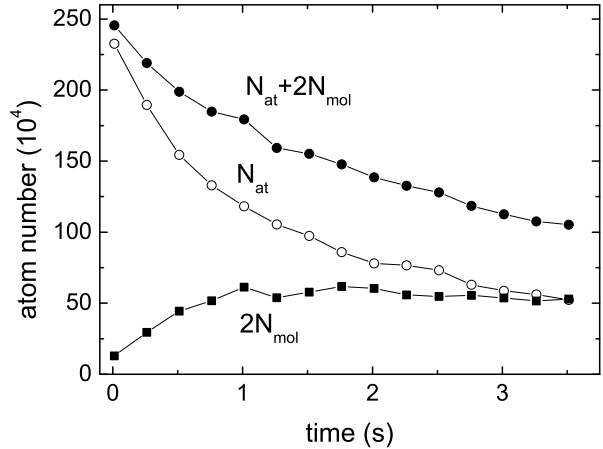}
\vspace{-3mm}
\end{center}
\caption{Experimental results \cite{Jochim2003pgo} demonstrating
how an ultracold \Li gas  approaches a chemical atom-molecule
equilibrium on the molecular side of the Feshbach resonance. The
experiment starts with a non-degenerate, purely atomic gas at a
temperature of $2.5\,\mu$K and a peak atomic phase-space density
of $0.04$. The magnetic field is set to 690\,G, where
$a=1420\,a_0$ and \Eb$/$\kB$= 15\,\mu$K. $N_{\rm at}$ and $N_{\rm
mol}$ denote the number of unbound atoms and the number of
molecules, respectively. The total number of unbound and bound
atoms $2 N_{\rm mol} + N_{\rm at}$ slowly decreases because of
residual inelastic loss, see \ref{ssec_dimers}.} \label{formation}
\end{figure}

We have experimentally studied the thermal atom-molecule
equilibrium in Ref.~\cite{Jochim2003pgo}. Fig.~\ref{formation}
illustrates how an initially pure atomic gas tends to an
atom-molecule equilibrium. The experiment was performed at a
magnetic field of 690\,G and a temperature $T = 2.5\,\mu$K with a
molecular binding energy of \Eb/\kB$= 15\,\mu$K. The observation
that more than 50\% of the atoms tend to form molecules at a
phase-space density of a factor of thirty from degeneracy,
highlights the role of the Boltzmann factor (see
Eq.~\ref{eq_equilibrium}) in the equilibrium. Note that in
Fig.~\ref{formation}, the total number of particles decreases
slowly because of residual inelastic decay of the molecules. The
magnetic field of 690\,G is too far away from resonance to obtain
a full suppression of inelastic collisions. Further experiments in
Ref.~\cite{Jochim2003pgo} also demonstrated how an atom-molecule
thermal equilibrium is approached from an initially pure molecular
sample. In this case atoms are produced through dissociation of
molecules at small molecular binding energies closer to the
Feshbach resonance.

An experiment at ENS \cite{Cubizolles2003pol} demonstrated the
adiabatic conversion of a degenerate \Li Fermi gas produced at
$a<0$ into a molecular gas by slowly sweeping across the Feshbach
resonance. This resulted in a large molecular fraction of up to
85\% and experimental conditions close to mBEC. Before the work in
$^6$Li, molecule formation in an ultracold Fermi gas through a
Feshbach sweep was demonstrated with \K at JILA
\cite{Regal2003cum}. The long-lived nature of the \K molecules
close to the Feshbach resonance was demonstrated in later work
\cite{Regal2004lom}. Note that long-lived molecules of \Li were
also produced from a degenerate gas at Rice Univ.\
\cite{Strecker2003coa}. This experiment, however, was performed by
sweeping across the narrow Feshbach resonance at 543\,G. The
observed stability cannot be explained in terms of the Pauli
suppression arguments in Sec.~\ref{ssec_dimers} and, to the best
knowledge of the author, still awaits a full interpretation.

\subsection{Evaporative cooling of an atom-molecule mixture}
\label{ssec_evaporation}

Based on the thermal atom-molecule equilibrium arguments discussed
before, we can now understand why the evaporation process works so
well on the molecular side of the Feshbach resonance.

Experimentally, we found that highly efficient evaporative cooling
can be performed at a fixed magnetic field around 764\,G
\cite{Jochim2003bec}. At this optimum field, the large scattering
length $a = +4500\,a_0$ warrants a large stability of the
molecules against inelastic decay (see \ref{ssec_dimers}). The
corresponding binding energy \Eb/\kB$= 1.5\,\mu$K is small enough
to minimize recombination heating during the cooling process.
However, it is larger than the typical Fermi energies in the final
evaporation stage of a few hundred nK, which favors the molecule
formation in the last stage of the cooling process.

\begin{figure}
\begin{center}
\includegraphics[width=13.5cm]{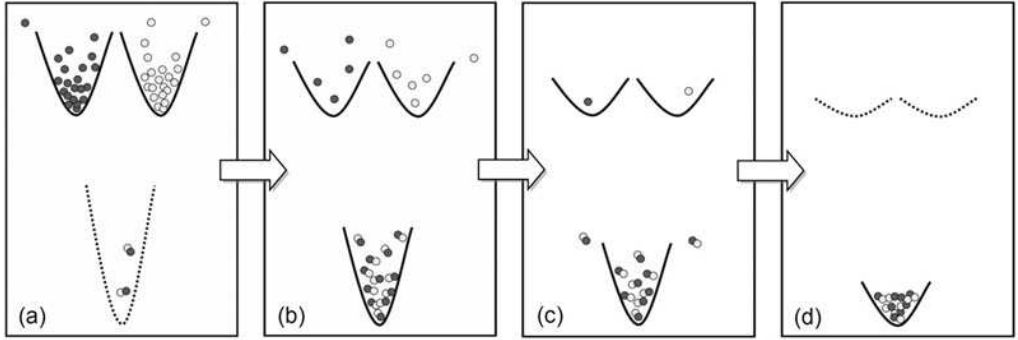}
\end{center}
\caption{Stages of evaporative cooling on the molecular side of
the Feshbach resonance. (a) For a hot gas, very few molecules are
present and the evaporation can be understood in terms of elastic
collisions in the atomic spin mixture. (b) As the gas gets colder
the chemical atom-molecule equilibrium begins to favor the
molecules. (c) Further evaporation removes atoms but not molecules
because of the two times different trap depths. (d) After
disappearance of the atoms, evaporation can be fully understood in
terms of the molecular gas. This eventually leads to molecular
Bose-Einstein condensation.} \label{evaporation}
\end{figure}

The different stages of evaporative cooling are illustrated in
Fig.~\ref{evaporation}. In the first stage (a) molecule formation
is negligible. As the cooling process proceeds (b, c), an
increasing part of the trapped sample consists of molecules. Here,
it is important to note that the optical trap is twice as deep for
the molecules. This is due to the weakly bound dimers having twice
the polarizability. Therefore, evaporation in an atom-molecule
mixture near thermal equilibrium essentially removes atoms and not
the molecules. This predominant evaporation of unpaired atoms also
has the interesting effect that the sample reaches a balanced
50/50 spin mixture, even if one starts the evaporation with some
imbalance in the spin composition~\footnote{A large initial
imbalance, however, is detrimental as the cooling process already
breaks down in the first stage where only atoms are present.}. In
the final stage of the evaporation process, only molecules are
left and the process can be essentially understood in terms of
elastic molecule-molecule interactions. This leads to the
formation of a molecular Bose-Einstein condensate (mBEC), as we
will discuss in more detail in Sec.~\ref{ssec_mBEC}

We point out two more facts to fully understand the efficiency of
the evaporative cooling process in our set-up. The magnetic field
that we use for Feshbach tuning of the scattering properties
\cite{JochimPhD} exhibits a curvature~\footnote{For technical
reasons the coils were not realized in the Helmholtz
configuration, where the curvature disappears. At the end this
turned out to be a lucky choice for the creation of the mBEC.},
which provides us with a magnetic trapping potential for the
high-field seeking atoms along the laser beam axis (corresponding
trapping frequency of $24.5\,{\rm Hz} \times \sqrt{B/{\rm kG}}$).
When the optical trap is very week at the end of the evaporation
process, the trap is a hybrid (optically for the transverse motion
and magnetically for the axial motion). The cooling then results
in an axial compression of the cloud which helps to maintain high
enough number densities. The second interesting fact, which makes
the all-optical route to degeneracy different for fermions and
bosons
\cite{Barrett2001aof,Weber2003bec,Rychtarik2004tdb,Kinoshita2005aob},
is that evaporative cooling of fermions can be performed at very
large scattering lengths. For bosons this is impossible because of
very fast three-body decay \cite{Roberts2000mfd,Weber2003bec}. For
a very large scattering length, a substantial part of the cooling
process proceeds in the unitarity limit, where the scattering
cross section is limited by the relative momentum of the
particles. Decreasing temperature leads to an increase in the
elastic scattering rate, which counteracts the effect of the
decreasing number density when the sample is decompressed. Axial
magnetic trapping and cooling in the unitarity limit help us to
maintain the high elastic collision rate needed for a fast cooling
process to degeneracy.

\begin{figure}
\begin{center}
\includegraphics[width=10cm]{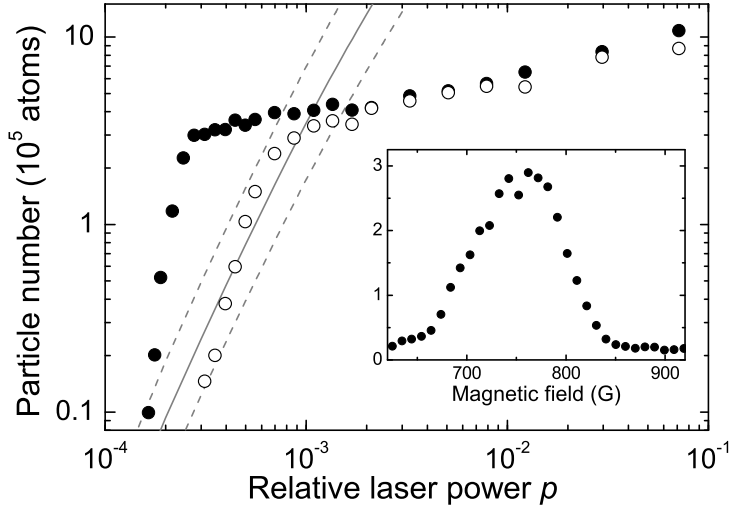}
\vspace{-3mm}
\end{center}
\caption{Evaporative cooling on both sides of the Feshbach
resonance exhibits a strikingly different behavior
\cite{Jochim2003bec}. The filled and open circles refer to
magnetic fields of 764\,G ($a=+4500a_0$) and 1176\,G
($a=-3000\,a_0$). We plot the total number of trapped particles $2
N_{\rm mol} + N_{\rm at}$ as a function of the laser power. The
power $p$ is given relative to the initial laser power of 10.5\,W
of an exponential evaporation ramp with a $1/e$ time constant of
230\,ms; the corresponding initial trap depth for the atoms is
$\sim$850$\mu$K. The solid line shows the maximum number of
trapped atoms according to the number of motional quantum states
of the trap. The dashed lines indicate the corresponding
uncertainty range due to the limited knowledge of the trap
parameters. The inset shows the optimum production of molecules in
the magnetic field range where a weakly bound molecular state
exists. Here the total number of particles is measured for various
magnetic fields at a fixed final ramp power $p = 2.8 \times
10^{-4}$, corresponding to a trap depth of $\sim$$440\,$nK for the
molecules.} \label{science2003}
\end{figure}

It is very interesting to compare evaporative cooling on the
molecular side of the Feshbach resonance ($a<0$) to the cooling on
the other side of the resonance ($a>0$). For similar values of
$|a|$, one obtains a comparable cross section $\sigma = 4\pi a^2$
for elastic collisions between atoms in the two spin states.
However, a striking difference shows up at low optical trap depth
in the final stage of the evaporative cooling process.
Fig.~\ref{science2003} shows how the number of trapped atoms
(including the ones bound to molecules) decreases with the trap
power. On the molecular side of the Feshbach resonance, a shallow
trap can contain about ten times more atoms than on the other side
of the resonance. Obviously, this cannot be understood in terms of
the scattering cross section of atoms and highlights a dramatic
dependence on the sign of the scattering length.

At the negative-$a$ side of the resonance (open symbols in
Fig.~\ref{science2003}) a sharp decrease of the number of trapped
particles is observed when the Fermi energy reaches the trap
depth. Lowering the trap power below this critical level leads to
a spilling of atoms out of the trap. The trapping potential does
simply not offer enough quantum states for the atoms. The observed
spilling is consistent with the number of quantum states
calculated for a non-interacting Fermi gas (solid line). A similar
spilling effect is observed at the molecular side of the resonance
($a>0$, filled symbols), but at much lower trap power. Before this
spilling sets in, the trap contains nearly ten times more atoms as
it would be possible for a non-interacting Fermi gas. This
striking effect is explained by the formation and Bose-Einstein
condensation of molecules. The spilling effect observed for the
molecules with decreasing trap depth shows the chemical potential
of the molecular condensate.

The strategy to evaporatively cool on the molecular side of the
Feshbach resonance and to produce an mBEC as the starting point
for further experiments is also followed at MIT, ENS, and Rice
University. The Duke group performs forced evaporation very close
to the resonance, which we believe to be a better strategy when
the dynamical range for the trap power reduction is technically
limited. Comparing the performance of evaporative cooling at
different magnetic fields, we observed that the cooling process is
somewhat more efficient and more robust on the molecular side of
the resonance than very close to resonance.

\subsection{The appearance of mBEC}
\label{ssec_mBEC}

At the time of our early mBEC experiments in fall 2003
\cite{Jochim2003bec} we had no imaging system to detect the
spatial distribution of the gas at high magnetic fields, where we
performed the evaporation experiments described in the preceding
section. Nevertheless, by measuring the dependence of the total
number of trapped particles on different parameters, we compiled
various pieces of evidence for the formation of mBEC:

\begin{enumerate}

\item We observed that a very shallow trap can contain much more
atoms than it offers quantum states for a weakly interacting
atomic Fermi gas.

\item We observed very long lifetimes of up to 40\,s for the
trapped sample after a fast and highly efficient evaporation
process. This shows that the sample has enough time to thermalize
into an equilibrium state.

\item We measured the frequency of a collective oscillation mode
(see also Sec.~\ref{sec_oscillations}), which clearly revealed
hydrodynamic behavior.

\item By controlled spilling of the quantum gas out of the trap
applying a variable magnetic gradient, we could demonstrate that
the chemical potential of the trapped sample depends on the
magnetic field in the way expected for a mBEC from the prediction
of the dimer-dimer scattering lenghts, see Eq.~\ref{eq_add}.


\end{enumerate}

\begin{figure}
\vspace{3mm}
\begin{center}
\includegraphics[width=12cm]{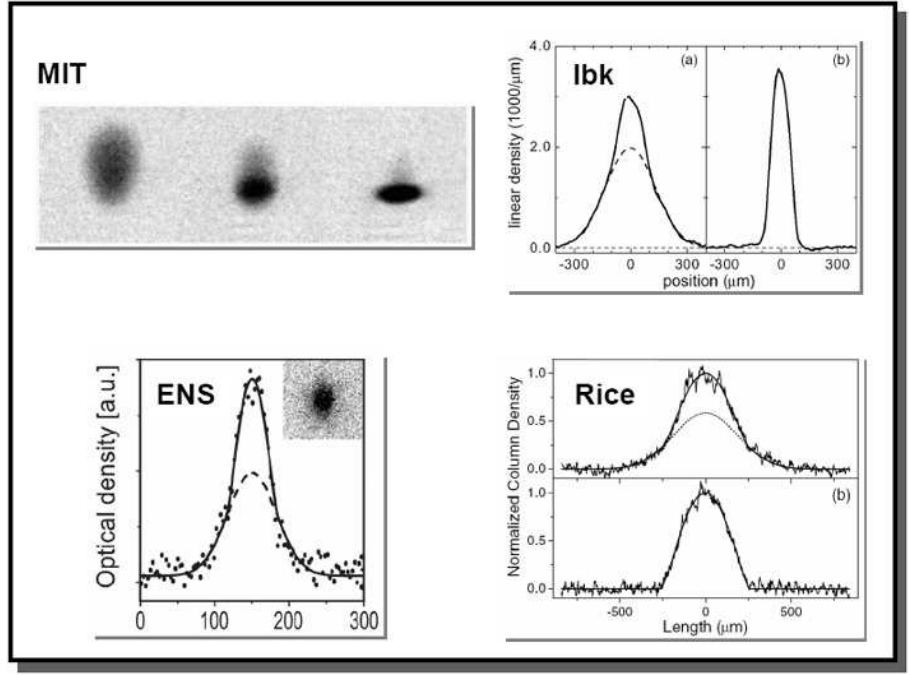}
\end{center}
\caption{Gallery of \Li molecular BEC experiments. Bimodal spatial
distributions were observed for expanding gases at MIT
\cite{Zwierlein2003oob} and at ENS \cite{Bourdel2004eso}, and in
{\em in-situ} profiles of the trapped cloud in Innsbruck
\cite{Bartenstein2004cfa} and at Rice University
\cite{Partridge2005mpo}.} \label{mBECgallery}
\end{figure}

These observations, together with our previous knowledge on
molecule formation in the gas \cite{Jochim2003pgo} and the general
properties of the weakly bound dimers \cite{Petrov2004wbm}, led us
to a consistent interpretation in terms of mBEC. At the same time
mBEC was observed in a \K gas at JILA in Boulder
\cite{Greiner2003eoa}. It is an amazing coincidence that our
manuscript was submitted for publication on exactly the same day
(Nov.\ 3, 2003) as the Boulder work on mBEC in \K. Very shortly
afterwards, the MIT group observed the formation of mBEC in \Li by
detecting bimodal spatial distributions of the gas expanding after
release from the trap \cite{Zwierlein2003oob}. A few weeks later,
we observed the appearance of bimodial distributions in {\em
in-situ} absorption images of the trapped cloud
\cite{Bartenstein2004cfa}. At about the same time also the ENS
group reported on mBEC. Fig.~\ref{mBECgallery} shows a gallery of
different observations of bimodal distributions in formation of
$^6$Li mBECs at MIT \cite{Zwierlein2003oob}, in Innsbruck
\cite{Bartenstein2004cfa}, at ENS \cite{Bourdel2004eso}, and at
Rice University \cite{Partridge2005mpo}.

\section{Crossover from mBEC to a fermionic superfluid}
\label{sec_crossover}

With the advent of ultracold Fermi gases with tunable interactions
a unique way has opened up to explore a long-standing problem in
many-body quantum physics, which has attracted considerable
attention since the seminal work by Eagles \cite{Eagles1969ppw},
Leggett \cite{Leggett1980xxx} and Nozi\`eres and Schmitt-Rink
\cite{Nozieres1985xxx}. Here we give a brief introduction
(\ref{ssec_crossoverbasics}) into the physics of the BEC-BCS
crossover~\footnote{In the condensed-matter literature, the
crossover is commonly referred to as the ``BCS-BEC crossover'',
because BCS theory served as the starting point. In our work on
ultracold gases, we use ``BEC-BCS crossover'', because we start
out with the molecular BEC. The physics is one and the same.} and
we introduce some basic definitions and typical experimental
parameters (\ref{ssec_definitions}). We then consider a universal
Fermi gas with resonant interactions (\ref{ssec_universal}) and
the equation of state in the crossover (\ref{ssec_eos}). Next we
discuss the crossover at non-zero temperatures, including the
isentropic conversion between different interaction regimes
(\ref{ssec_isentropic}). We finally review our first crossover
experiments where we have observed how spatial profiles and the
size of the strongly interacting, trapped cloud changed with
variable interaction strength (\ref{ssec_sizemeas}).

\begin{figure}
\vspace{3mm}
\begin{center}
\includegraphics{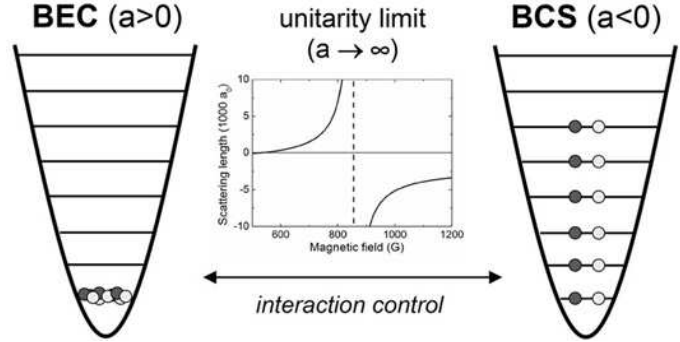}
\end{center}
\vspace{-3mm} \caption{Illustration of the BEC-BCS crossover in a
zero-temperature \Li Fermi gas with tunable interactions. For
positive scattering length ($a>0$, BEC side of the Feshbach
resonance) the ground state of the system is a Bose-Einstein
condensate of molecules. On resonance ($a\rightarrow \pm \infty$,
unitarity limit) a strongly interacting Fermi gas with universal
properties is realized. For negative scattering length ($a<0$, BCS
side of the resonance) the system approaches the BCS regime.}
\label{crossover}
\end{figure}

\subsection{BEC-BCS crossover physics: a brief introduction}
\label{ssec_crossoverbasics}

The crossover of a superfluid system from the BEC regime into the
Bardeen-Cooper-Schrieffer (BCS) regime  can be intuitively
understood by first considering the two limits, which can be
described in the framework of well-established theory (see
illustration in Fig.~\ref{crossover}). For moderate positive
scattering lengths, the fermions form bosonic molecules, and the
ground state at $T=0$ is a BEC. For moderate negative scattering
lengths, the ground state at $T=0$ is the well-known BCS state
\cite{Bardeen1957tos,Tinkham1996book}. With variable interaction
strength across a resonance, both regimes are smoothly connected
through the strongly interacting regime. Here both BEC and BCS
approaches break down and the description of the strongly
interacting system is a difficult task. This situation poses great
challenges for many-body quantum theories \cite{Chen2005bbc}.

The nature of pairing is the key to understanding how the system
changes through the crossover. On the BEC side, the pairs are
molecules which can be understood in the framework of two-body
physics. The molecular binding energy \Eb is large compared with
all other energies, and the molecules are small compared with the
typical interparticle spacing. In this case, the interaction can
be simply described in terms of molecule-molecule collisions, for
which the scattering length is known (see \ref{ssec_dimers}). On
the BCS side, two atoms with opposite momentum form Cooper pairs
on the surface of the Fermi sphere. The pairing energy, i.e.\ the
``pairing gap'', is small compared with the Fermi energy \EF and
the Cooper pairs are large objects with a size greatly exceeding
the typical interparticle spacing. In the strongly interacting
regime, the pairs are no longer pure molecules or Cooper pairs.
Their binding energy is comparable to the Fermi energy and their
size is about the interparticle spacing. One may consider them
either as generalized molecules, stabilized by many-body effects,
or alternatively as generalized Cooper pairs.

The ground state at $T=0$ is a superfluid throughout the whole
crossover. In the BEC limit, the fermionic degrees of freedom are
irrelevant and superfluidity can be fully understood in terms of
the bosonic nature of the system. In the opposite limit,
superfluidity is described in the framework of BCS theory
\cite{Bardeen1957tos,Tinkham1996book}. In the strongly interacting
regime, a novel type of superfluidity (``resonance superfluidity''
\cite{Holland2001rsi,Timmermans2001poc}) occurs where both bosonic
and fermionic degrees of freedom are important .

\subsection{Basic definitions, typical experimental parameters}
\label{ssec_definitions}

Let us start with some basic definitions, which we will need to
describe the physics in the rest of this contribution. The {\em
Fermi energy} of a trapped, non-interacting two-component gas is
given by
\begin{equation}\label{eq_defEF}
    E_{\rm F} = \hbar \bar{\omega} \, (3N)^{1/3}\,,
\end{equation}
here $N$ is the total number of atoms in both spin states, and
$\bar{\omega} = (\omega_{\rm x}\omega_{\rm y}\omega_{\rm
z})^{1/3}$ is the geometrically averaged oscillation frequency in
the harmonic trapping potential. This expression can be derived
within the Thomas-Fermi approximation for a sufficiently large
number of trapped atoms \cite{Butts1997tfg}. The chemical
potential $\mu$ of the non-interacting gas is equal to \EF.

We use the Fermi energy of a non-interacting gas \EF to define an
{\em energy scale} for the whole BEC-BCS crossover, i.e.\ for any
regime of interactions. The corresponding Fermi temperature is
\begin{equation}\label{eq_defTF}
   T_{\rm F} = E_{\rm F}/k_{\rm B} \,.
\end{equation}

We now introduce a Fermi wave number $k_{\rm F}$, following the
relation
\begin{equation}\label{eq_defkF}
    \frac{\hbar^2 k_{\rm F}^2}{2m} = E_{\rm F}\,.
\end{equation}
The inverse Fermi wave number $k^{-1}_{\rm F}$ defines a typical
{\em length scale} for the crossover problem. For the
non-interacting case, \kF is related to the peak number density
$n_0$ in the center of the trap \cite{Butts1997tfg} by
\begin{equation}
n_0= \frac{k^{3}_{\rm F}}{3\pi^2} \,.
\end{equation}

To characterize the interaction regime, we introduce the
dimensionless interaction parameter \kFa, which is commonly used
to discuss crossover physics. We can now easily distinguish
between three different regimes. The BEC regime is realized for
\kFa$\gg 1$, whereas the BCS regime is obtained for \kFa$\ll -1$.
The strongly interacting regime lies between these two limits
where $1/k_{\rm F}|a|$ being small or not greatly exceeding unity.

Let us consider typical experimental parameters for our \Li spin
mixture: an atom number $N$ of a few $10^5$, and a mean trap
frequency $\bar{\omega}/2\pi$ near 200\,Hz. This corresponds to a
typical Fermi temperature $T_{\rm F} \approx 1\,\mu$K and to
$k^{-1}_{\rm F} \approx 200\,{\rm nm} \approx 4000\,a_0$.  A
comparison of $k_{\rm F}^{-1}$ with the scattering lengths close
to the 834\,G Feshbach resonance (see Fig.~\ref{resonance}) shows
that there is a broad crossover region where the \Li system is
strongly interacting. The peak number then considerably exceeds
the typical value $n_0 \approx 4 \times 10^{12}$\,cm$^{-3}$
calculated for the non-interacting case.

\subsection{Universal Fermi gas in the unitarity limit}
\label{ssec_universal}

The resonance where $a(B)$ diverges and \kFa$=0$, is at the heart
of BEC-BCS crossover physics~\footnote{In nuclear physics this
situation is known as the ``Bertsch problem''. G.~F.~Bertsch
raised the question on the ground state properties of neutron
matter under conditions where the scattering length between the
two neutron spin states is large compared to the interparticle
spacing.}. Here the $s$-wave interaction between colliding
fermions is as strong as quantum mechanics allows within the
fundamental limit of unitarity. In this situation, \EF and
$1/k_{\rm F}$ represent the only energy and length scales in the
problem and the system acquires universal properties
\cite{Baker1999nmm,Heiselberg2001fsw,Ho2004uto}. The broad
Feshbach resonance in the ultracold \Li gas offers excellent
possibilities to study the properties of the universal Fermi gas
\cite{Bruun2004uoa} and the situation has attracted a great deal
of experimental interest, as described in various parts of these
proceedings.


At $T=0$, universality implies a simple scaling behavior with
respect to the situation of a non-interacting Fermi gas. Following
the arguments in
Refs.~\cite{Baker1999nmm,Heiselberg2001fsw,Ohara2002ooa} the
atomic mass $m$ can be simply replaced by an effective mass
\begin{equation}
m_{\rm eff} = \frac{m}{1+\beta}\,, \label{eq_effmass}
\end{equation}
where $\beta \simeq -0.57$ \cite{Carlson2003sfg,
Astrakharchik2004eos} is a dimensionless, universal many-body
parameter. For a harmonic trapping potential, Eq.~\ref{eq_effmass}
results in an effective trap frequency $\bar{\omega}_{\rm eff} =
\sqrt{1+\beta}\,\bar{\omega}$, and the chemical potential for
zero-temperature gas in the unitarity limit is then given by
\begin{equation}
\mu = \sqrt{1+\beta}\,E_{\rm F}\,. \label{eq_mubeta}
\end{equation}
The density profile of the universal Fermi gas with resonant
interactions is just a simple rescaled version of the density
profile of the non-interacting gas, smaller by a factor of
$(1+\beta)^{1/4} \simeq 0.81$.

The universal many-body parameter was recently calculated based on
quantum Monte-Carlo methods, yielding $\beta = -0.56(1)$
\cite{Carlson2003sfg} and $-0.58(1)$ \cite{Astrakharchik2004eos}.
A diagrammatic theoretical approach \cite{Perali2004qcb} gave a
value $-0.545$ very close to these numerical results. Several
experiments in \Li \cite{Bartenstein2004cfa,
Bourdel2004eso,Kinast2005hco,Partridge2005pap} and in \K
\cite{Stewart2006peo} have provided measurements of $\beta$ in
good agreement with the theoretical predictions. We will discuss
our experimental results on $\beta$ in some more detail in
\ref{ssec_sizemeas}.

At $T \ne 0$, the Fermi gas with unitarity-limited interactions
obeys a universal thermodynamics with $T/T_{\rm F}$ being the
relevant dimensionless temperature parameter \cite{Ho2004uto}.
Thermodynamic properties of the system have been experimentally
studied in Ref.~\cite{Kinast2005hco}.

\subsection{Equation of state}
\label{ssec_eos}

The equation of state is of central interest to characterize the
interaction properties of the Fermi gas in the BEC-BCS crossover.
For a system at $T=0$, the equation of state is described by the
chemical potential $\mu$ as a function of the number density $n$.
For $\mu(n)$ at $T=0$, we now consider three special cases. For a
{non-interacting Fermi gas}, $\mu = E_{\rm F}$, and one thus
obtains
\begin{equation}\label{eq_mu1}
\mu  =  (3\pi^2)^{2/3} \frac{\hbar^2}{2 m} \, n^{2/3}\,.
\end{equation}
For a Fermi gas with resonant interactions, universality implies
that one obtains the same expression with a prefactor $1+\beta$
(Eq.~\ref{eq_effmass}).
In the mBEC regime, the chemical potential for the dimers is
$\mu_{\rm d} = {4\pi\hbar^2 a_{\rm dd}}{m^{-1}_{\rm d}}\,n_{\rm
d}$. With the simple relations between mass ($m_{\rm d} = 2m$),
number density ($n_{\rm d} = n/2$), and scattering length ($a_{\rm
dd} = 0.6a$, see Eq.~\ref{eq_add}) for dimers and atoms, and after
substraction of the molecular binding energy \Eb$=
\hbar^2/(ma^2)$, we obtain
\begin{equation}\label{eq_mu3}
\mu = \frac{1}{2}\,(\mu_{\rm d} - E_{\rm b}) = \frac{\hbar^2}{2m}
\left(0.6 \pi \, a \, n - a^{-2}\right)\,.
\end{equation}

For the general BEC-BCS crossover problem one can introduce a
``polytropic'' equation of state \cite{Heiselberg2004cmo} in the
form
\begin{equation}
\mu \propto n^{\gamma} \, .
\label{eq_eos}
\end{equation}
Here the ``polytropic index'' $\gamma$ depends on the interaction
parameter \kFa. By comparing this equation of state with the above
expressions one immediately sees that $\gamma=1$ for the mBEC case
(\kFa$\gg 1$), $\gamma = 2/3$ both for the unitarity limit (\kFa$=
0$) and for the non-interacting case (\kFa$\ll -1$). These three
values are fixed boundary conditions for any crossover theory
describing $\gamma$ as a function of \kFa.

In the experiments, the Fermi gases are usually confined in nearly
harmonic trapping potentials, which leads to an inhomogeneous
density distribution. If the trap is not too small one can
introduce the {\em local-density approximation} and consider a
local chemical potential $\mu({\bf r}) = \mu - U({\bf r})$, which
includes the trapping potential $U({\bf r})$ at the position ${\bf
r}$. This assumption holds if the energy quantization of the trap
is irrelevant with respect to the chemical potential and the
pair-size is small compared to the finite size of the trapped
sample. This approximation is well fulfilled for all crossover
experiments performed in Innsbruck.

\subsection{Phase-diagram, relevant temperatures and energies}
\label{ssec_isentropic}

At finite temperatures the BEC-BCS crossover problem becomes very
challenging  and it is of fundamental interest to understand the
phase diagram of the gas. Two temperatures play an important role,
the temperature $T_{\rm c}$ for the superfluid phase transition
and a pairing temperature $T^*$, characterizing the onset of
pairing. Let us first discuss these two temperatures in the three
limits of the crossover (BEC, unitarity, and BCS), see first two
rows in Table~\ref{tab_crossover}.

\begin{table}
\caption{Overview of important temperatures
and energies
in the three crossover
limits. The expressions are valid for a harmonically trapped Fermi
gas.} \label{tab_crossover}
\begin{tabular}{|lc|c|c|c|}
  \hline
  && mBEC & unitarity & BCS \\
   && ($1/k_{\rm F}a \gg 1$) & ($1/k_{\rm F}a = 0$) & ($1/k_{\rm F}a \ll -1$) \\
 \hline
 crit.\ temp.\ & $T_{\rm c}$ &  $0.518\,T_{\rm F}$ &
$\sim0.3\,T_{\rm F}$ & $0.277 \, T_{\rm F}
\exp\left(\frac{\pi}{2k_{\rm
  F}a}\right)$ \\
\smallskip  pair.\ temp.\ & $T^*$   & $12 \left( \frac{T}{T_{\rm F}}\right)^3 = \exp \left(\frac{T_F}{T}\,\frac{2}{(k_{\rm F}a)^{2}}\right)$ & $\sim0.4\,T_{\rm F}$ & $T_{\rm c}$ \\
\smallskip  gap energy  & $2 \Delta$              & $2 \, (k_{\rm F}a)^{-2} E_{\rm F}$ & $1.8 E_{\rm F}$ &  $3.528 \, k_{\rm B} T_{\rm c}$ \\
\smallskip  chem.\ pot.\  & $\mu$              & $\left( 0.294 (k_{\rm F} a)^{2/5} - (k_{\rm F} a)^{-2} \right) E_{\rm F} $ & $0.66 E_{\rm F}$ &  $E_{\rm F}$ \\
  \hline
\end{tabular}
\end{table}

The critical temperature $T_{\rm c}$ in the mBEC limit follows
directly from the usual expression for the BEC transition
temperature in a harmonic trap $k_{\rm B} T_{\rm c} \simeq 0.94
\hbar \bar{\omega} N^{1/3}_{\rm m}$ \cite{Dalfovo1999tob}, $N_{\rm
m}=N/2$, and $k_{\rm B} T_{\rm F} = \hbar \bar{\omega}
(3N)^{1/3}$. The given value for the critical temperature in the
unitarity limit was derived in various crossover theories
\cite{Perali2004bbc,Kinast2005hco}. For the BCS regime, the
critical temperature is a well-known result from
Ref.~\cite{Gorkov1961xxx}; see also
Refs.~\cite{Combescot1999tla,Carr2004aab}.

For the pairing temperature $T^*$, typical numbers are given in
the second row of Table~\ref{tab_crossover}. In the framework of
BCS theory, there is no difference between $T^*$ and $T_{\rm c}$,
which means that as soon as Cooper pairs are formed the system is
also superfluid. On the BEC side, however, molecules are formed at
much higher temperatures as the phase transition to molecular BEC
occurs (see discussion in \ref{ssec_equilibrium}). Setting
$\phi_{\rm mol} = \phi_{\rm at}$ in Eq.~\ref{eq_equilibrium}, one
can derive the implicit equation for $T^*/T_{\rm F}$ given in the
Table. In the unitarity limit, $T^{*}$ is not much higher than
$T_{\rm c}$; Ref.~\cite{Perali2004bbc} suggests $T^{*}/T_{\rm c}
\approx 1.3$.

\begin{figure}
\vspace{3mm}
\begin{center}
\includegraphics{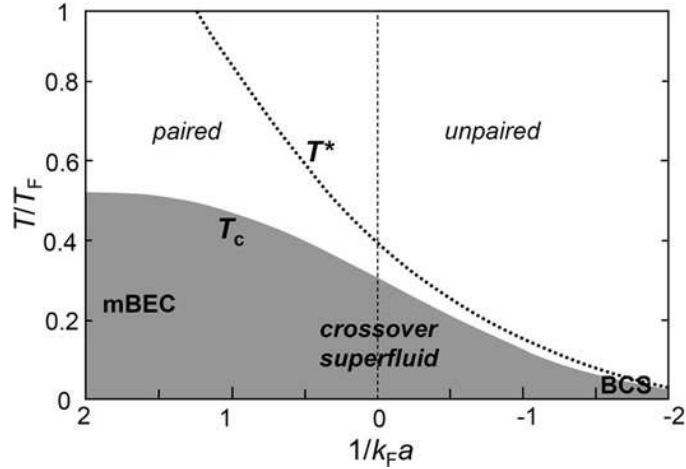}
\end{center}
\vspace{-3mm} \caption{Schematic phase diagram for the BEC-BCS
crossover in a harmonic trapping potential \cite{Perali2004bbc}.
The critical temperature $T_{\rm c}$ marks the phase transition
from the normal to the superfluid phase. Pair formation sets in
gradually at a typical temperature $T^* > T_{\rm c}$.}
 \label{phasediag}
\end{figure}

The phase diagram in Fig.~\ref{phasediag} illustrates the behavior
of $T_{\rm c}$ and $T^*$, as discussed before for the three
limits. We point out that, in strongly interacting Fermi gases,
there is a certain region where pairing occurs without
superfluidity. In the language of high-T$_{\rm c}$
superconductivity \cite{Stajic2004nos,Chen2005bbc}, ``preformed
pairs'' are present in the ``pseudo-gap regime''.

For overview purposes, Table~\ref{tab_crossover} also gives the
pairing energy $2\Delta$ and the chemical potential $\mu$. In the
mBEC regime, the pairing energy just corresponds to the molecular
binding energy $E_{\rm b} = 2 (k_{\rm F}a)^{-2}E_{\rm F}$. The
chemical potential $\mu = \frac{1}{2}(\mu_{\rm d} - E_{\rm b})$
(see Eq.~\ref{eq_mu3}) can be derived from $\mu_{\rm
d}/\hbar\bar{\omega} = \frac{1}{2}(15N_{\rm d} a_{\rm d}/a_{\rm
ho})^{2/5}$ with $a_{\rm ho} = (\hbar/m_{\rm
d}\bar{\omega})^{1/2}$, valid for an mBEC in the Thomas-Fermi
limit \cite{Dalfovo1999tob}. In the BCS limit, there is the fixed
relation of the ``gap'' $\Delta$ to the critical temperature
$T_{\rm c}$ given in Table \ref{tab_crossover}. For the unitarity
limit, the value given for the pairing energy stems from quantum
Monte-Carlo calculations \cite{Carlson2003sfg}. The behavior of
$\Delta$ in the crossover is extensively discussed in
\ref{ssec_gap_crossover}. The table also presents the chemical
potential $\mu$ according to Eqs.~\ref{eq_mu1} and \ref{eq_mu3},
rewritten in terms of the parameters $1/k_{\rm F}a$ and $E_{\rm
F}$.

\subsection{First Innsbruck crossover experiments: conservation of entropy, spatial profiles, and potential energy of the trapped gas}
\label{ssec_sizemeas}

The possibility to continuously vary the interaction parameter
\kFa through Feshbach tuning  offers the fascinating possibility
to convert the Fermi gas between different regimes and thus to
explore the BEC-BCS crossover. We performed our first experiments
on crossover physics \cite{Bartenstein2004cfa} in December 2003
shortly after the first creation of the mBEC \cite{Jochim2003bec}.
Here we summarize the main results of these early experiments,
which are of general importance for BEC-BCS crossover experiments
with \Li.

We performed slow conversion-reconversion cycles, in which the
strongly interacting gas was adiabatically converted from the BEC
side of the crossover to the BCS side and vice versa. We found
that this conversion took place in a lossless way and that the
spatial profiles of the trapped cloud did not show any significant
heating. We could thus demonstrate that, under appropriate
experimental conditions, the conversion process can proceed in an
essentially {\em adiabatic and reversible} way, which means that
the {\em entropy} of the gas is conserved.

The conservation of entropy has important consequences for the
experiments: Because of the different relations between entropy
and temperature in various interaction regimes, an isentropic
conversion in general changes the temperature. As a substantial
benefit, a drastic temperature reduction occurs when the
degenerate gas is converted from mBEC into the BCS regime. This is
very favorable  for the achievement of a superfluid state on the
BCS side of the resonance \cite{Carr2004aab} or in the
unitarity-limited resonance regime \cite{Chen2005toi}. In our
experiments, we typically start out with a condensate fraction of
more than 90\% in the weakly interacting mBEC regime. Based on the
isentropic conversion process and the thermodynamics discussed in
Ref.~\cite{Chen2005toi} we estimate that we obtain typical
temperatures between $0.05\,T_{\rm F}$ and $0.1\,T_{\rm F}$ for
the Fermi gas in the unitarity limit~\footnote{We note that the
large stability of \Li in the mBEC regime offers an advantage over
\K in that one can evaporatively cool in the mBEC regime and
exploit the temperature-reduction effect in conversion onto the
BCS side of the resonance.}.

\begin{figure}
\vspace{2mm}
\begin{center}
\includegraphics{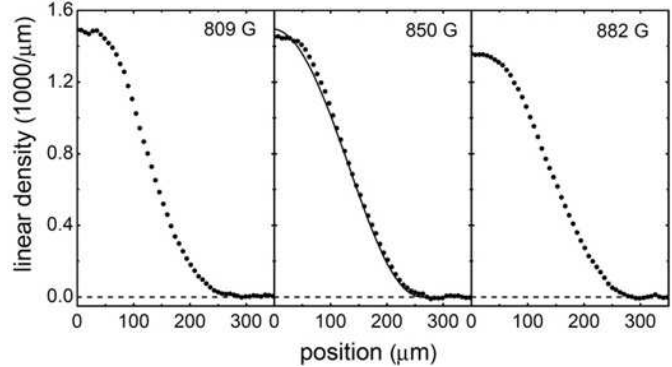}
\end{center}
\vspace{-3mm} \caption{Axial density profiles of a trapped \Li
Fermi gas in the crossover region \cite{Bartenstein2004cfa}. The
middle profile, taken very close to resonance (850\,G), is
compared to the Thomas-Fermi profile of a universal Fermi gas
(solid line). The small deviation on the top is due to a residual
interference pattern in the images.} \label{axialprofiles}
\end{figure}

Using slow magnetic field ramps we isentropically converted the
trapped gas into different interaction regimes covering the whole
resonance region and beyond. By {\em in-situ} imaging we recorded
the axial density profiles of the trapped cloud. The results (see
profiles in Fig.~\ref{axialprofiles}) demonstrated the smooth
behavior in the crossover. The cloud just became larger without
showing any particular features, and we found simple Thomas-Fermi
profiles to fit our observations very well. The one-dimensional
spatial profiles did not show any signatures of a superfluid phase
transition~\footnote{This is different in an imbalanced
spin-mixture, where the superfluid phase transition was observed
by changes in the spatial profiles \cite{Zwierlein2006doo}.}, in
agreement with theoretical expectations
\cite{Perali2004bbc,Stajic2005dpo}.

\begin{figure}
\vspace{0mm}
\begin{center}
\includegraphics{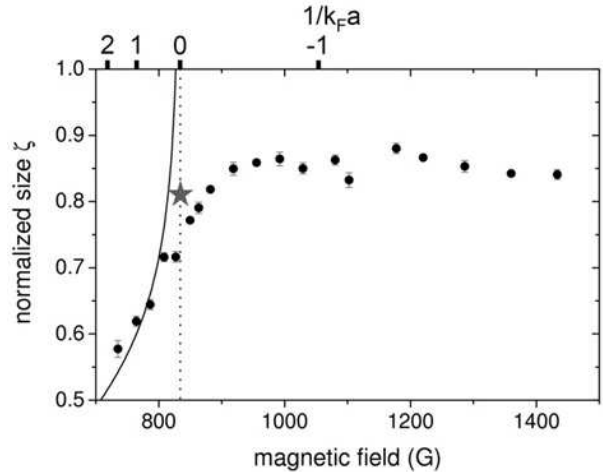}
\end{center}
\vspace{-3mm} \caption{Results of the first Innsbruck BEC-BCS
crossover experiments \cite{Bartenstein2004cfa} on the axial size
normalized to the theoretical size of a non-interacting Fermi gas
($\zeta = z_{\rm rms}/z_0$). The solid line is a theoretical
prediction for the size of a mBEC in the Thomas-Fermi limit, and
the star indicates the theoretical value for the unitarity limit.}
\label{axialsize}
\end{figure}

To quantitatively characterize the behavior, we measured the
root-mean-square axial size $z_{\rm rms}$ of the cloud as a
function of the magnetic field $B$. The normalized quantity $\zeta
= z_{\rm rms}/z_0$ gives the relative size as compared to a
non-interacting Fermi gas, where $z_0 = (E_{\rm F}/4m\omega^2_{\rm
z})^{1/2}$. The {\em potential energy} of the harmonically trapped
gas relative to a non-interacting Fermi gas is then simply given
by $\zeta^2$. Within the local density approximation, this is also
valid for the three-dimensional situation. Our experimental
results can thus be interpreted as the first measurements of the
potential energy of a trapped Fermi gas near $T=0$ in the BEC-BCS
crossover~\footnote{We note that a later thorough analysis of the
conditions of the experiments in Ref.~\cite{Bartenstein2004cfa}
confirmed the atom number $N=4\times10^5$ to within an uncertainty
of $\pm$30\%. However, we found that the horizontal trap frequency
was only 80\% of the value that we used based on the assumption of
a cylindrically symmetric trap (see \ref{ssec_breakdown}).
Moreover, the exact position of the Feshbach resonance was located
at 834\,G (see \ref{ssec_tunability}) instead of 850\,G as assumed
in the first analysis of the experiment. The up-to-date values are
used for Fig.~\ref{axialsize}, causing slight deviations from the
original presentation of our data.}.

The measured values for the relative size $\zeta$ are plotted in
Fig.~\ref{axialsize} and compared to the predictions for a weakly
interacting molecular BEC with $a_{\rm dd}= 0.6 a$ in the in the
Thomas-Fermi limit (solid line) and a universal Fermi gas in the
unitarity limit (star). The experimental data on the mBEC side are
consistent with the theoretical prediction. On resonance, the
measured size was found somewhat below the prediction
$(1+\beta)^{1/4} \simeq 0.81$ (\ref{ssec_universal}); see star in
Fig.~\ref{axialsize}. This slight discrepancy, however, may be
explained by possible calibration errors in the measured number of
atoms and in the magnification of the imaging system in
combination with the anharmonicity of the radial trapping
potential. Beyond resonance the results stayed well below the
non-interacting value $\zeta=1$, showing that we did not reach
weakly interacting conditions, This is a general consequence of
the large background scattering length of \Li, which (in contrast
to \K) makes it very difficult to realize a weakly interacting
Fermi gas on the BCS side of the Feshbach resonance.

In general, the dependence of the size and thus the potential
energy of the trapped gas in the BEC-BCS crossover that we
observed in our first experiments \cite{Bartenstein2004cfa} was
found to fit well to corresponding theoretical predictions
\cite{Perali2004qcb}. Later experiments by other groups provided
more accurate measurements on the size of the gas for the
particulary interesting unitarity limit
\cite{Kinast2005hco,Partridge2005pap,Stewart2006peo}.

\section{Collective excitations in the BEC-BCS crossover}
\label{sec_oscillations}

Elementary excitation modes provide fundamental insight into the
properties of quan\-tum-dege\-ne\-rate gases. In particular, they
provide unique experimental access to study the hydrodynamic
behavior that is associated with superfluidity.
Collective modes have been studied very early in atomic BEC
research, both in experiments \cite{Jin1996ceo,Mewes1996ceo} and
in theory \cite{Stringari1996ceo}. Measurements on collective
oscillations have proven powerful tools for the investigation of
various phenomena in atomic BECs
\cite{Jin1997tdd,Stamperkurn1998cah,Onofrio2000seo,Marago2000oot,Chevy2002tbm}.
Building on this rich experience, collective modes attracted
immediate attention to study strongly interacting Fermi gases
\cite{Stringari2004coo,Kinast2004efs,Bartenstein2004ceo} as soon
as these systems became experimentally available. Here, we give a
basic introduction into collective modes in the BEC-BCS crossover
(\ref{ssec_modebasics}), and we present an overview of the major
experimental results obtained in our laboratory in Innsbruck and
at Duke University (\ref{ssec_modeoverview}), before we discuss
our results in some more detail
(\ref{ssec_axialmode}--\ref{ssec_modenew}).

\subsection{Basics of collective modes}
\label{ssec_modebasics}

\begin{figure}
\vspace{3mm}
\begin{center}
\includegraphics{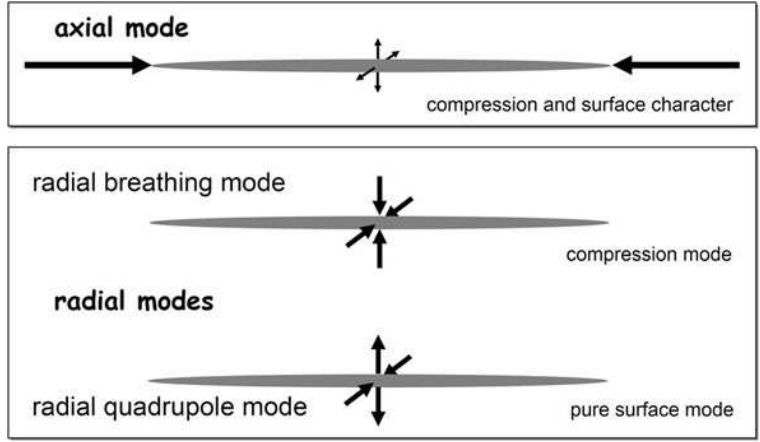}
\end{center}
\vspace{-1mm}
 \caption{Illustration of elementary collective modes
of a cigar-shaped quantum gas, confined in an elongated trap. The
axial mode corresponds to a slow oscillation with both compression
and surface character. The two low-lying radial modes correspond
to fast oscillations with strong compression character (``radial
breathing mode'') and with pure surface character (``radial
quadrupole mode'').} \label{collmodes}
\end{figure}

We will focus our discussion on the geometry of elongated traps
with cylindrical symmetry, because this is the relevant geometry
for strongly interacting Fermi gases in single-beam optical traps.
Besides the simple sloshing modes that correspond to
center-of-mass oscillations in the trap, the cigar-shaped quantum
gas exhibits three elementary, low-lying collective modes, which
are illustrated in Fig.~\ref{collmodes}.

The {\em axial mode} corresponds to an oscillation of the length
of the ``cigar'' with a frequency of the order of the axial trap
frequency $\omega_{\rm z}$. This oscillation is accompanied by a
$180^{\circ}$ phase-shifted oscillation of the cigar's radius,
which reflects a quadrupolar character of the mode. Thus, the mode
has the mixed character of a compression and a surface mode. The
frequencies of the two low-lying {\em radial modes} are of the
order of the radial trap frequency $\omega_{\rm r}$. The ``radial
breathing mode'' is a compression mode, for which the radius of
the sample oscillates. The ``radial quadrupole mode'' is a pure
surface mode where a transverse deformation oscillates without any
change of the volume.

To understand collective modes in a Fermi gas, it is crucial to
distinguish between two fundamentally different regimes. Which
regime is realized in an experiment depends on the interaction
strength \kFa and the temperature $T$ of the gas.

\begin{itemize}

\item {\em The collisionsless, non-superfluid regime} -- In a
weakly interacting degenerate Fermi gas, elastic collisions are
Pauli blocked \cite{Demarco2001pbo,Ferrari1999cri,Gupta2004ciz}.
This is due to the fact that the final states for elastic
scattering processes are already occupied. This Pauli blocking
effect has dramatic consequences for the dynamics of a
two-component Fermi gas when it is cooled down to degeneracy. In
the non-degenerate case, the influence of collisions between the
two different spin states can be very strong, as it is highlighted
by our efficient evaporative cooling process (see
\ref{ssec_equilibrium}). In the degenerate case, however,
collisions are strongly suppressed. A substantial increase in
relaxation times \cite{Vichi2000cdo} shows up as an important
consequence.

\item {\em The hydrodynamic regime} -- When a superfluid is formed
at sufficiently low temperatures, hydrodynamic behavior occurs as
an intrinsic property of the system, and the gas follows the
equations of {\em superfluid hydrodynamics} (see lecture of
S.~Stringari in these proceedings). However, in a strongly
interacting Fermi gas bosonic pairs can be formed and their
elastic interactions are no longer Pauli blocked; this may lead to
{\em classical hydrodynamics} in a degenerate gas. In this case,
the sample follows basically the same hydrodynamic equations as in
the superfluid case. Therefore, it is not possible to draw an
immediate conclusion on superfluidity just from the observation of
hydrodynamic behavior.

\end{itemize}

The existence of these two different regimes has important
consequences for collective oscillations. In the (non-superfluid)
collisionless case, the fermionic atoms perform independent
oscillations in the trapping potential and the effect of elastic
collisions and collisional relaxation is small
\cite{Vichi1999coo,Vichi2000cdo}. The ensemble then shows
decoupled oscillations along the different degrees of freedoms
with frequencies that are twice the respective trap frequencies
\footnote{We neglect small interaction shifts, which are discussed
in \cite{Vichi1999coo}.}.

In the hydrodynamic regime, a solution of the equations of motion
(see lecture of S.~Stringari) yields the following expressions for
the collective mode frequencies in the elongated trap limit,
$\omega_{\rm z}/\omega_{\rm r} \rightarrow 0$~\footnote{For all
experiments reported here, the traps fulfilled $\omega_{\rm
z}/\omega_{\rm r} < 0.1$, which makes the elongated trap limit a
valid approximation.}:
\begin{eqnarray}
\omega_{\rm ax} & = & \sqrt{(3\bar{\gamma}+2)/(\bar{\gamma}+1)} \,\, \omega_{\rm z}\,,  \label{eq_axmodefreq}\\
\omega_{\rm c} & = & \sqrt{2\bar{\gamma} + 2} \,\, \omega_{\rm r}\,,  \label{eq_cmodefreq}\\
\omega_{\rm q} & = & \sqrt{2} \,\, \omega_{\rm r}\,.
\label{eq_qmodefreq}
\end{eqnarray}
Here $\bar{\gamma}$ is an effective polytropic index for the
equation of state (Eq.~\ref{eq_eos}), which takes into account the
variation of the density across the inhomogeneous sample in the
harmonic trap \cite{Hu2004cma,Astrakharchik2005eos}. For the mBEC
case $\bar{\gamma} = 1$, and for the unitarity limit $\bar{\gamma}
= 2/3$. The theory of collective modes in the BEC-BCS crossover
has attracted considerable interest and is extensively discussed
in
Refs.~\cite{Stringari2004coo,Heiselberg2004cmo,Hu2004cma,Kim2004tdd,Combescot2004ace,
Bulgac2005coo,Manini2005bac,Astrakharchik2005eos,Desilva2005coo,Combescot2006cmo}.

\begin{table}
\caption{Overview of collective mode frequencies in different
regimes.} \label{tab_freq}
\begin{tabular}{|rr|c|c|c|}
  \hline
  && \multicolumn{2}{c|}{\em hydrodynamic} & {\em collisionless,} \\
  && mBEC & unitarity & {\em non-superfluid} \\
   && ($1/k_{\rm F}a \gg 1$) & ($1/k_{\rm F}a = 0$) & \\
  \hline
   axial mode & $\omega_{\rm ax}/\omega_{\rm z}$ & $\sqrt{5/2}=1.581..$ & $\sqrt{12/5} = 1.549..$ & 2 \\
 radial compression mode & $\omega_{\rm c}/\omega_{\rm r}$ & 2 & $\sqrt{10/3}=1.826..$ & 2 \\
  radial quadrupole mode  & $\omega_{\rm q}/\omega_{\rm r}$ & $\sqrt{2}=1.414..$ & $\sqrt{2}=1.414..$ & 2 \\
  \hline
\end{tabular}
\end{table}

Table \ref{tab_freq} presents an overview of the frequencies of
the three low-lying modes in different regimes. When the
interaction is varied from mBEC to the unitarity limit, the axial
mode changes its frequency by just $\sim$2\%. However, for the
radial breathing mode the relative change is five times larger
($\sim$10\%). This difference reflects the much stronger
compression character of the radial breathing mode, which is why
this mode is a prime tool to experimentally investigate the
equation of state (see \ref{ssec_precisecoll}). The fact that the
radial quadrupole mode is a pure surface mode makes it insensitive
to the equation of state. This mode can thus serve as a powerful
tool for investigating the large differences between hydrodynamic
and collisionless behavior \cite{Altmeyer2006q}.

\subsection{Overview of recent experiments}
\label{ssec_modeoverview}

Here we give a brief overview of the major results of collective
mode experiments performed at Duke University and in Innsbruck.
Already in our early work on mBEC \cite{Jochim2003bec} we measured
the axial mode frequency to show that the trapped sample behaved
hydrodynamically. The first experimental results on collective
modes in the BEC-BCS crossover were reported by our team and the
Duke group at the {\it Workshop on Ultracold Fermi Gases} in
Levico (4-6 March 2004). These results were published in
Refs.~\cite{Kinast2004efs,Bartenstein2004ceo}.

\begin{figure}
\vspace{0mm}
\begin{center}
\includegraphics{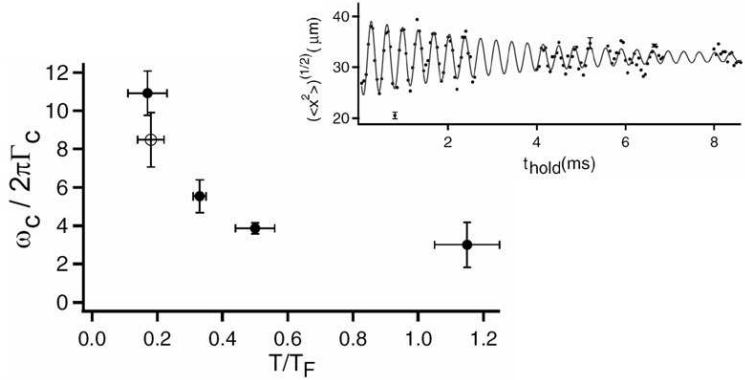}
\end{center}
\vspace{-3mm} \caption{Evidence for superfluidity in a strongly
interacting Fermi gas obtained at Duke University from
measurements of the damping of the radial breathing mode
\cite{Kinast2004efs}. The
 damping time $1/\Gamma_c$ (in units of the oscillation period $2\pi/\omega_c$)
is plotted versus temperature $T$. The inset shows a breathing
oscillation ($\omega_{\rm c}/2\pi = 2830$\,Hz) at the lowest
temperatures reached in the experiment. This figure was adapted
from Ref.~\cite{Kinast2004efs}.} \label{duke_evidence}
\end{figure}

The Duke group investigated the radial breathing mode for resonant
interactions and measured a frequency that was consistent with the
theoretical prediction (see \ref{ssec_modebasics}) for a
hydrodynamic Fermi gas with unitarity-limited interactions. They
also investigated the temp\-erature-dependent damping behavior and
observed strongly increasing damping times when the sample was
cooled well below the Fermi temperature \TF; the main result is
shown in Fig.~\ref{duke_evidence}. By comparing the results with
available theories on Fermi gases in the collisionless,
non-superfluid regime and with theories on collisional
hydrodynamics they found the observed behavior to be inconsistent
with these two regimes \cite{Kinast2004efs}. Superfluidity
provided a plausible explanation for these observations, and the
Duke group thus interpreted the results as {\it evidence for
superfluidity}. Later experiments
\cite{Chin2004oop,Kinast2005hco,Zwierlein2005vas} indeed provided
a consistent picture of superfluidity for the conditions under
which these collective mode experiments were performed.

In our early experiments \cite{Bartenstein2004ceo}, we measured
the frequencies of the axial mode and the radial compression mode
in the BEC-BCS crossover. Here we observed the frequency
variations that result from the changing equation of state. On the
BCS side of the Feshbach resonance, we observed a transition from
hydrodynamic to non-superfluid, collisonless behavior. The
transition occurred rather smoothly in the axial mode (see
\ref{ssec_axialmode}) but abruptly in the radial breathing mode
(see \ref{ssec_breakdown}). We also observed ultralow damping in
the axial mode, which nicely fits into the picture of
superfluidity. The abrupt breakdown of hydrodynamics in the radial
breathing mode was also observed at Duke University
\cite{Kinast2004boh}.

Further experiments on collective modes at Duke University
\cite{Kinast2005doa} provided more information on the temperature
dependence of damping for unitarity-limited interactions. This
experiment also hinted on different damping regimes. At Innsbruck
University, we carried out a series of precision measurements on
the frequencies of collective modes in the crossover
\cite{Altmeyer2006pmo}. This provided a test of the equation of
state and resolved seeming discrepancies between state-of-the-art
theoretical predictions \cite{Astrakharchik2005eos} and the early
experiments \cite{Bartenstein2004ceo,Kinast2004boh}.

\subsection{Axial mode}
\label{ssec_axialmode}

\begin{figure}
\vspace{0mm}
\begin{center}
\includegraphics{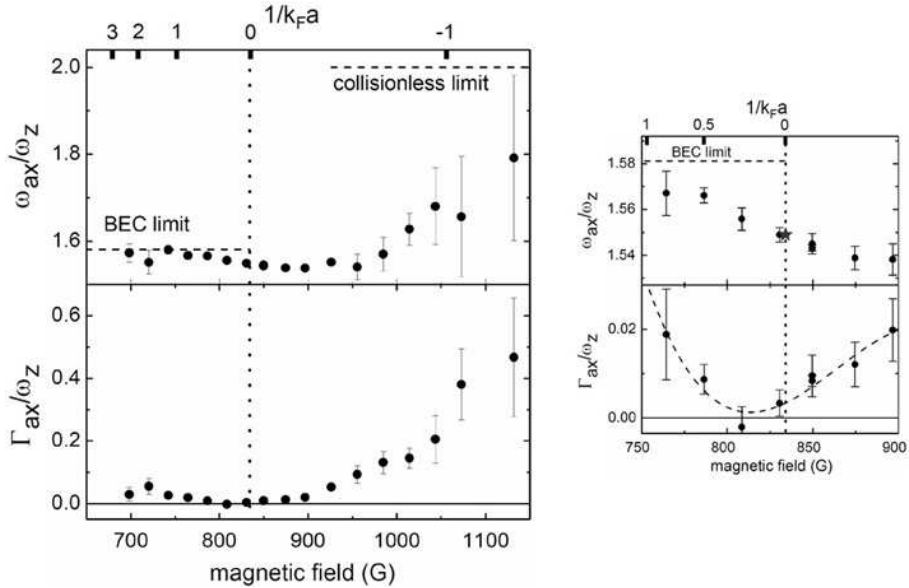}
\end{center}
\vspace{-3mm} \caption{Axial mode in the BEC-BCS crossover. The
figure shows our measurements \cite{Bartenstein2004ceo} of the
frequency $\omega_{\rm ax}$ and the damping rate $\Gamma_{\rm ax}$
in units of the axial trap frequency $\omega_{\rm z}$
($\omega_{\rm z}/2\pi = 22.6$\,Hz at $B=834\,$G, $\omega_{\rm
r}/2\pi \approx 700$\,Hz). The horizontal, dashed lines indicate
the theoretically expected frequencies in the BEC limit and in the
collisionless limit (cf.\
Table~\ref{tab_freq}). 
The figure on the right-hand
side shows a blow-up of the resonance region; here the star refers
to the frequency expected for the unitarity limit.}
\label{axialmode}
\end{figure}

Our measurements of frequency and damping of the axial mode
\cite{Bartenstein2004ceo} are shown in Fig.~\ref{axialmode}. To
tune the two-body interaction we varied the magnetic field in a
range between 700 and 1150\,G, corresponding to a variation of the
interaction parameter \kFa between $2.5$ and $-1.2$. For magnetic
fields up to $\sim$900\,G (\kFa$\approx -0.45$), the oscillation
shows the hydrodynamic frequencies and very low damping. For
higher fields, damping strongly increases and the frequency gets
closer to the collisionless value, but never reaches it
completely. These observations are consistent with a gradual
transition from hydrodynamic to  collisionless behavior
\cite{Vichi2000cdo}. Even far on the BCS side of the resonance,
the true collisionless regime is not reached, as the Pauli
blocking effect is not strong enough to suppress elastic
collisions on a time scale below the very long axial oscillation
period of about $50\,$ms.

At the right-hand side of Fig.~\ref{axialmode}, we show a blow-up
of the resonance region. One clearly sees that the axial mode
frequency changes from the BEC value $\omega_{\rm ax}/\omega_{\rm
z} = \sqrt{5/2} = 1.581$ to the value of a universal Fermi gas in
the unitarity limit of $\sqrt{12/5} = 1.549$. We were able to
detect this small 2\% effect because of the very low damping of
the mode, allowing long observation times. Moreover, the magnetic
axial confinement was perfectly harmonic, and the corresponding
trap frequency $\omega_{\rm z}$ could be measured with a relative
uncertainty of below $10^{-3}$ \cite{BartensteinPhD}.

It is also very interesting to consider the damping of the axial
mode. The minimum damping rate was observed at $\sim$815\,G, which
is slightly below the exact resonance (834\,G). Here we measured
the very low value of $\Gamma_{\rm z}/\omega_{\rm z} \approx
0.0015$, which corresponds to a $1/e$ damping time as large as
$\sim$5\,s. According to our present knowledge of the system, this
ultralow damping is a result of superfluidity of the strongly
interacting Fermi gas. The damping observed for lower magnetic
fields can be understood as a consequence of heating due to
inelastic processes in the gas~\footnote{Precise frequency
measurements of the slow axial mode require long observation
times. On the BEC side of the resonance, heating due to inelastic
decay then becomes a hardly avoidable problem.}. In general,
damping rates are very sensitive to the residual temperature of
the sample.

\subsection{Radial breathing mode: breakdown of hydrodynamics}
\label{ssec_breakdown}

\begin{figure}
\vspace{0mm}
\begin{center}
\includegraphics{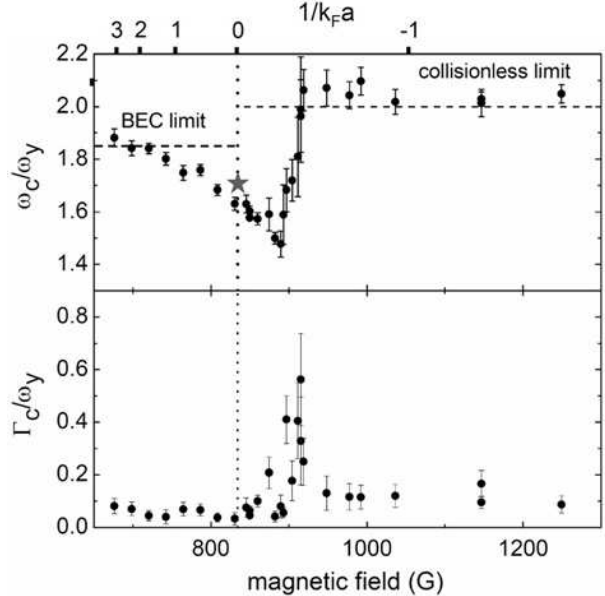}
\end{center}
\vspace{-3mm} \caption{Measurements of the frequency $\omega_{\rm
c}$ and the damping rate $\Gamma_{\rm c}$ of the radial
compression mode in the BEC-BCS crossover
\cite{Bartenstein2004ceo}. Here the oscillation frequencies were
determined relative to the vertical trap frequency $\omega_{\rm y}
\approx 750$\,Hz. As we found in later experiments, the trap was
somewhat elliptic with a horizontal trap frequency $\omega_{\rm x}
\approx 600$\,Hz. This is about 20\% below the vertical one and
has a substantial effect on the hydrodynamic frequencies
\cite{Altmeyer2006noc}. On the BEC side of the Feshbach resonance,
the dashed line indicates the frequency theoretically expected in
the BEC limit ($\omega_{\rm c}/\omega_{\rm y}=1.85$ for
$\omega_{\rm x}/\omega_{\rm y}=0.8$). The star marks the frequency
in the unitarity limit ($\omega_{\rm c}/\omega_{\rm y}=1.805$ for
$\omega_{\rm x}/\omega_{\rm y}=0.8$). On the BCS-side, the dashed
line indicates the frequency $2\omega_{\rm y}$ for the
collisionless case.}
\label{radialmode}
\end{figure}

Our early measurements \cite{Bartenstein2004ceo} of frequency and
damping of the radial breathing mode are shown in
Fig.~\ref{radialmode}. The most striking feature is a sharp
transition from the hydrodynamic to the collisionless regime. This
occurs at a magnetic field of $\sim$900\,G (\kFa$\approx -0.45$).
Apparently, the hydrodynamic regime extends from the mBEC region
across the unitarity limit onto the BCS side of the resonance.
This behavior is consistent with the direct observation of
superfluidity in the crossover region through vortices
\cite{Zwierlein2005vas}. The breakdown of superfluid hydrodynamics
is accompanied by very fast damping, which indicates a fast
dissipation mechanism in the sample. We will come back to this in
our discussion of the pairing gap (last paragraph in
\ref{ssec_gap_crossover}). The breakdown of hydrodynamics on the
BCS side of the resonance was also observed by the Duke group
\cite{Kinast2004boh}.

To quantitatively understand the frequencies in the hydrodynamic
regime as measured in our early collective mode experiments
\cite{Bartenstein2004ceo}, one has to take into account an
unintended ellipticity of the transverse trapping potential
\cite{Altmeyer2006noc}. We found out later after technical
upgrades to our apparatus that the ratio of horizontal and
vertical trap frequencies was $\omega_{\rm x}/\omega_{\rm y}
\approx 0.8$. Due to the fact that the gas was not completely
cylindrically symmetric, the collective mode frequencies deviate
from the simple expressions presented in \ref{ssec_modebasics}.
For small ellipticities, Eq.~\ref{eq_cmodefreq} still provides a
reasonable approximation when an effective transverse oscillation
frequency of $\sqrt{\omega_{\rm x}\omega_{\rm y}}$ is used for
$\omega_{\rm r}$ \cite{Kinast2004efs}. For an accurate
interpretation of the measurements, however, a more careful
consideration of ellipticity effects is necessary
\cite{Thomas2005vta,Altmeyer2006pmo,Altmeyer2006noc}.

In Fig.~\ref{radialmode}, we indicate the expected normalized
compression mode frequencies $\omega_{\rm c}/\omega_{\rm y}$ for
the limits of mBEC (dahed line below resonance), unitarity (star),
and for a non-interacting collisionless gas (dashed line above
resonance). For normalization we have used the vertical trap
frequency $\omega_{\rm y}$, which was directly measured in the
experiments. Moreover, we assumed $\omega_{\rm x}/\omega_{\rm y} =
0.8$ to calculate the eigenfrequencies of the collective modes
\cite{Altmeyer2006noc}. We see that, within the experimental
uncertainties, the measurements agree reasonably well with those
limits~\footnote{The slight deviation in the unitarity limit is
likely due to the anharmonicity of the trapping potential, which
is not taken into account in the calculation of the frequencies.}.

For a quantitative comparison of our early compression mode
measurement with theory and also with the experiments of the Duke
group, the ellipticity turned out to be the main problem. However,
as an unintended benefit of this experimental imperfection, the
larger difference between the frequencies in the hydrodynamic and
the collisionsless regime strongly enhanced the visibility of the
transition between these two regimes.

\subsection{Precision test of the equation of state}
\label{ssec_precisecoll}

Collective modes with compression character can serve as sensitive
probes to test the equation of state of a superfluid gas in the
BEC-BCS crossover. The fact that a compression mode frequency is
generally lower for a Fermi gas in the unitarity limit than in the
mBEC case simply reflects the larger compressibility of a Fermi
gas as compared to a BEC. The data provided by the experiments in
Innsbruck \cite{Bartenstein2004ceo} and at Duke University
\cite{Kinast2004efs,Kinast2004boh} in 2004 opened up an intriguing
possibility for quantitative tests of BEC-BCS crossover physics.
For such precision tests, frequency measurements of collective
modes are superior to the simple size measurements discussed in
Sec.~\ref{sec_crossover}. It is an important lesson that one
learns from metrology that it is often very advantageous to
convert the quantity to be measured into a frequency. In this
spirit, the radial breathing mode can be seen as an excellent
instrument to convert compressibility into a frequency for
accurate measurements.

A comparison of the axial mode data from Innsbruck and the
measurements on the radial breathing mode from Duke University
with mean-field BCS theory showed reasonable agreement
\cite{Hu2004cma,Manini2005bac,Astrakharchik2005eos}. This,
however, was somewhat surprising as mean-field BCS theory has the
obvious shortcoming that it does not account for beyond-mean-field
effects \cite{Lee1957mbp,Lee1957eae}. The latter were expected to
up-shift the compression mode frequencies in the strongly
interacting mBEC regime \cite{Stringari2004coo}, but they seemed
to be absent in the experiments. Advanced theoretical calculations
based on a quantum Monte-Carlo approach
\cite{Astrakharchik2004eos} confirmed the expectation of
beyond-mean-field effects in the equation of state and
corresponding up-shifts in the collective mode frequencies as
compared to mean-field BCS theory
\cite{Manini2005bac,Astrakharchik2005eos}.

\begin{figure}
\vspace{3mm}
\begin{center}
\includegraphics[width=8cm]{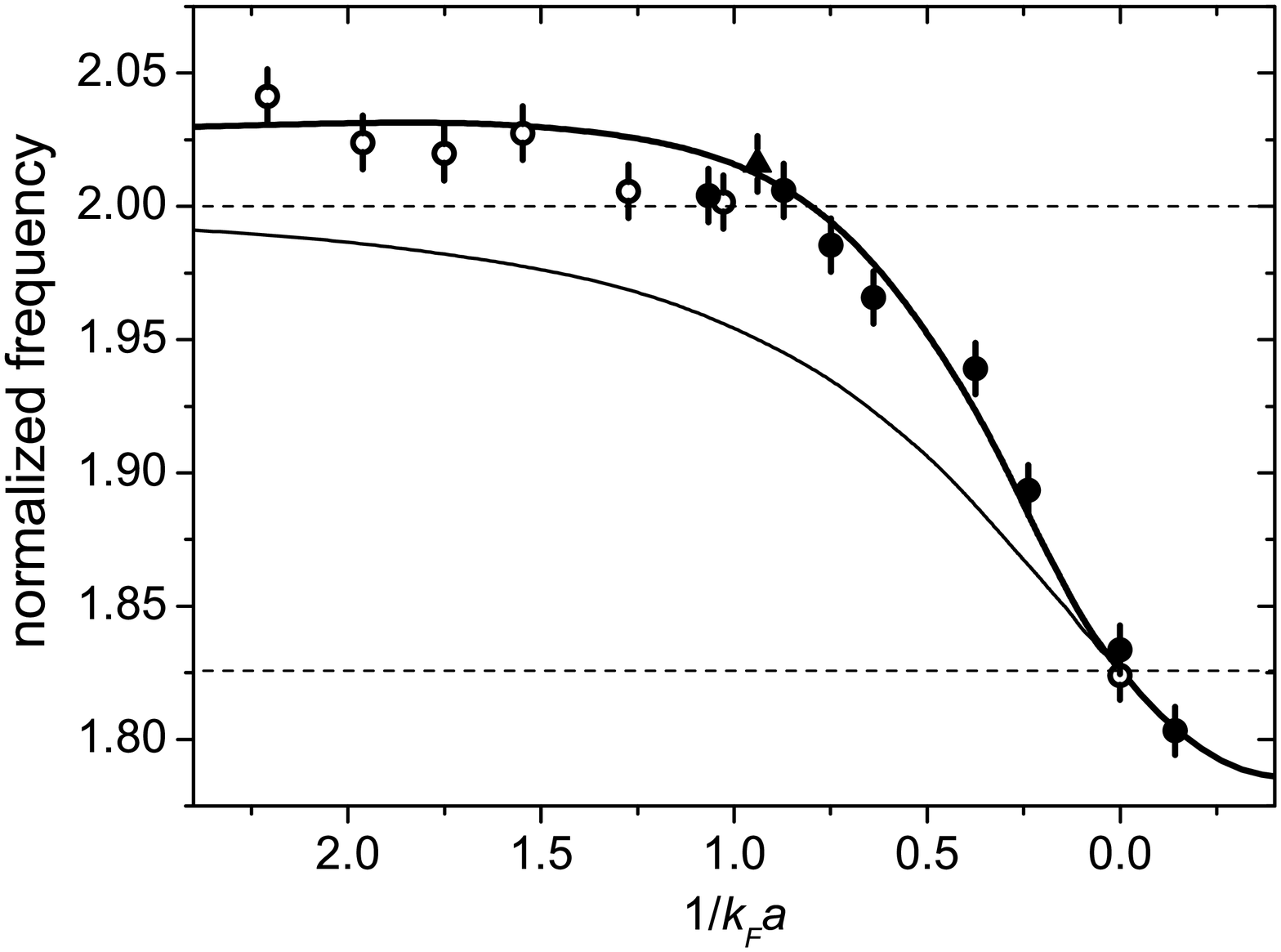}
\end{center}
\vspace{-3mm} \caption{Precision measurements of the radial
breathing mode frequency versus interaction parameter \kFa in
comparison with theoretical calculations \cite{Altmeyer2006pmo}.
The frequency is normalized to the radial trap frequency. The
experimental data include small corrections for trap ellipticity
and anharmonicity and can thus be directly compared to theory in
the limit of an elongated trap with cylindrical symmetry (see
\ref{ssec_modebasics}). The open and filled circles refer to
measurements at trap frequencies $(\omega_{\rm x}\omega_{\rm
y})^{1/2}$ of 290\,Hz and 590\,Hz, respectively. Here $\omega_{\rm
x}/\omega_{\rm y}$ was typically between $0.91$ and $0.94$. The
filled triangle shows a zero-temperature extrapolation of a set of
measurements on the temperature dependence of the frequency. The
theory curves refer to mean-field BCS theory (lower curve) and
quantum Monte-Carlo calculations (upper curve) and correspond to
the data presented in Ref.~\cite{Astrakharchik2005eos}. The
horizontal dashed lines indicate the values for the BEC limit and
the unitarity limit (see Table~\ref{tab_freq}).}
\label{precisecoll}
\end{figure}

The apparent discrepancy between theory and experiments
\footnote{The discrepancy between the first experiments at Duke
\cite{Kinast2004efs} and in Innsbruck \cite{Bartenstein2004ceo}
disappeared when we understood the problem of ellipticity in our
setup (see Sec.~\ref{ssec_breakdown}).} motivated us to perform a
new generation of collective mode experiments
\cite{Altmeyer2006pmo} with much higher precision and with much
better control of systematic effects. To achieve a $10^{-3}$
accuracy level, small ellipticity and anharmonicity corrections
had to be taken into account. In Fig.~\ref{precisecoll} we present
our measurements on the frequency of the radial breathing mode in
the BEC-BCS crossover. Because of the very low uncertainties it
can be clearly seen that our data agrees with the quantum
Monte-Carlo equation of state, thus ruling out mean-field BCS
theory. Our experimental results also demonstrate the presence of
the long-sought beyond-mean-field effects in the strongly
interacting BEC regime, which shift the normalized frequency
somewhat above the value of two, which one would obtain for a
weakly interacting BEC.

To obtain experimental results valid for the zero-temperature
limit (Fig.~\ref{precisecoll}) it was crucial to optimize the
timing sequence to prepare the gas in the BEC-BCS crossover with a
minimum of heating after the production of the mBEC as a starting
point. A comparison of the ultralow damping rates observed in our
new measurements with the previous data from 2004 shows that the
new experiments were indeed performed at much lower temperatures.
We are convinced that temperature-induced shifts provide a
plausible explanation for the earlier measurements being closer to
the predictions of mean-field BCS theory than to the more advanced
quantum Monte-Carlo results. For the strongly interacting mBEC
regime, we indeed observed heating (presumably due to inelastic
processes) to cause significant down-shifts of the breathing mode
frequency \cite{Altmeyer2006pmo}.

To put these results into a broader perspective, our precision
measurements on collective modes in the BEC-BCS crossover show
that ultracold Fermi gases provide a unique testing ground for
advanced many-body theories for strongly interacting systems.

\subsection{Other modes of interest}
\label{ssec_modenew}

\begin{figure}
\vspace{3mm}
\begin{center}
\includegraphics{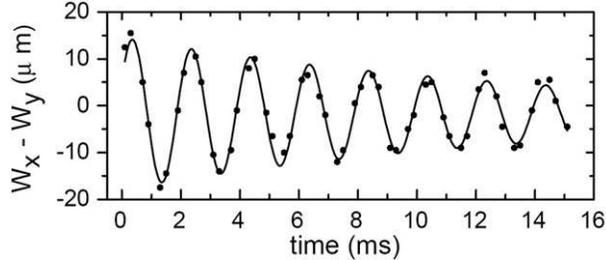}
\end{center}
\vspace{-5mm} \caption{Radial quadrupole oscillation of the
universal Fermi gas with unitarity-limited interactions at
$B=834\,$G. We plot the difference in horizontal and vertical
widths after a free expansion time of 2\,ms as a function of the
variable hold time in the trap. The measured oscillation frequency
$\omega_{\rm q}/2\pi = 499$\,Hz exactly corresponds to
$\sqrt{2}\,\omega_{\rm r}$.} \label{q_oscill}
\end{figure}

At the time of the Varenna Summer School we had started a set of
measurements on the radial quadrupole mode in the BEC-BCS
crossover \cite{Altmeyer2006q}. This mode had not been
investigated before. We implemented a two-dimensional
acousto-optical scanning system for the trapping laser beam; this
allows us to produce time-averaged optical potentials
\cite{Milner2001obf,Friedman2001ooc}, in particular potentials
with variable ellipticities. With this new system, it is
straightforward to create an appropriate deformation of the
trapping potential to excite the quadrupole mode and other
interesting modes. Here we just show the oscillation of the radial
quadrupole mode for a universal Fermi gas right on resonance at
$B=834$\,G for the lowest temperatures that we can achieve
(Fig.~\ref{q_oscill}). The mode indeed exhibits the expected
frequency ($\omega_{\rm q} = \sqrt{2}\,\omega_{\rm r}$), which
nicely demonstrates the hydrodynamic behavior. Moreover, we find
that the damping is considerably faster than for the radial
compression mode at the same temperature.

Scissors modes
\cite{Gueryodelin1999sma,Marago2000oot,Minguzzi2001smi,Cozzini2003smo}
represent another interesting class of collective excitations
which we can investigate with our new system. A scissors mode is
excited by a sudden rotation of an elliptic trapping potential.
Scissors modes may serve as a new tool to study the temperature
dependence of hydrodynamics in the BEC-BCS crossover. Scissors
modes are closely related to rotations \cite{Cozzini2003fgi} and
may thus provide additional insight into the collisional or
superfluid nature of hydrodynamics.

\section{Pairing-gap spectroscopy in the BEC-BCS crossover}
\label{sec_gap}

In the preceding sections we have discussed our experiments on
important {\em macroscopic} properties of the strongly interacting
Fermi gas, like potential energy, hydrodynamics, and the equation
of state. We will now present our experimental results on the
observation of the ``pairing gap'' \cite{Chin2004oop}, which is a
{\em microscopic} property essential in the context of
superfluidity. The gap shows the pairing energy and thus
characterizes the nature of pairing in the crossover; see
discussion in \ref{ssec_crossoverbasics}.

Historically, the observation of a pairing gap marked an important
experimental breakthrough in research on superconductivity in the
1950s \cite{Biondi1956mwa,Glover1956tos,Tinkham1996book}. The gap
measurements provided a key to investigating the paired nature of
the particles responsible for the frictionless current in metals
at very low temperatures. The ground-breaking BCS theory
\cite{Bardeen1957tos,Tinkham1996book}, developed at about the same
time, showed that two electrons in the degenerate Fermi sea can be
coupled by an effectively attractive interaction and form a
delocalized, composite particle with bosonic character. BCS theory
predicted that the gap in the low-temperature limit is
proportional to the critical temperature $T_c$ (Eq.~\ref{TBCS}),
which was in agreement with the experimental observations from gap
spectroscopy.


Here we will first discuss radio-frequency spectroscopy as our
method to investigate pairing in different regimes
(\ref{ssec_gap_rfspec}). We will then show how molecular pairing
can be investigated and precise data on the binding energy can be
obtained (\ref{ssec_gap_mols}). Finally, we will discuss our
results on pairing in the many-body regimes of the crossover
(\ref{ssec_gap_crossover}), including the temperature dependence
of the gap.

\subsection{Basics of radio-frequency spectroscopy}
\label{ssec_gap_rfspec}

Radio-frequency (RF) spectroscopy has proven a powerful tool for
investigating interactions in ultracold Fermi gases. In 2003, the
method was introduced by the JILA group for \K
\cite{Regal2003mop,Regal2003cum} and by the MIT group for \Li
\cite{Gupta2003rfo}.

\begin{figure}
\vspace{3mm}
\begin{center}
\includegraphics{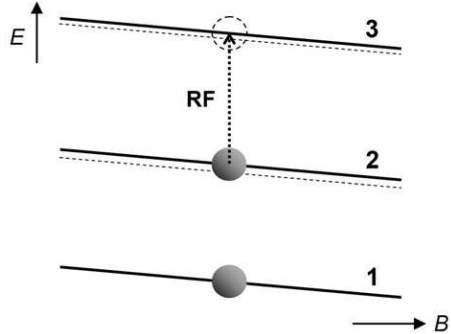}
\end{center}
\vspace{-3mm} \caption{Illustration of the basic principle of RF
spectroscopy for \Li at high-magnetic fields (see also
Fig.~\ref{paschenback}). The three states $1$, $2$, and $3$
essentially differ by the orientation of the nuclear spin ($m_I =
1$, $0$, $-1$, respectively). By driving RF transitions the spins
can be flipped and atoms are transferred from state $2$ to the
empty state $3$. In the region of the broad Feshbach resonance,
the splitting between the RF coupled states $2$ and $3$ is about
82\,MHz. The RF does not couple states $1$ and $2$ because of
their smaller splitting of about 76\,MHz. In the mean-field
regime, interactions result in effective level shifts (dashed
lines).} \label{rfspectro}
\end{figure}

The basic idea of RF spectroscopy can be easily understood by
looking on the simple Zeeman diagram of \Li in the high-field
region; see also section \ref{ssec_paschenback} for a more
detailed discussion of the energy levels. The \Li spin mixture
populates states 1 and 2 (magnetic quantum numbers $m_I=1, 0$),
whereas state 3 ($m_I = -1$) is empty. RF-induced transitions
can transfer atoms
from state 2 to the empty state 3. The experimental signature in a
state-selective detection scheme (e.g.\ absorption imaging) is the
appearance of particles in state 3 or the disappearance of
particles in state 2~\footnote{In a dense gas of \Li, atoms in
state 3 show a very rapid decay, which we attribute to three-body
collisions with 1 and 2. With atoms in three different spin
states, a three-body recombination event is not Pauli suppressed
and therefore very fast. This is the reason why all our
measurements show the loss from state 2 instead of atoms appearing
in 3.}.

In the non-interacting case, the transition frequency is
determined by the magnetic field through the well-known Breit-Rabi
formula. We found that interactions are in general very small for
``high'' temperatures of a few \TF. We perform such measurements
for the calibration of the magnetic field used for interaction
tuning. In the experiment, the transition frequency can be
determined within an uncertainty of $\sim$100\,Hz, which
corresponds to magnetic field uncertainties as low as a
$\sim$20\,mG.

In the mean-field regime of a weakly interacting Fermi gas,
interactions lead to a shift of the RF transition frequency
\cite{Regal2003mop} given by $\Delta \nu_{\rm mf} = 2\hbar
m^{-1}\,n_1 (a' - a)$, where $n_1=n/2$ is the number density of
atoms in state $1$ and $a'$ is the scattering length for
interactions between atoms in 1 and 3. In experimental work on \K
\cite{Regal2003mop}, this mean-field shift was used to measure the
change of the scattering length $a$ near a Feshbach resonance
under conditions where $a'$ just gives a constant, non-resonant
offset value. In \Li the interpretation of the differential
mean-field shift is somewhat more complicated because both
scattering length $a$ and $a'$ show resonant behavior
\cite{Gupta2003rfo,Bartenstein2005pdo}.

In the strongly interacting regime, the MIT group made the
striking observation of the absence of interaction shifts
\cite{Gupta2003rfo}. This experimental finding, which is also of
great relevance for the interpretation of our results on RF
spectroscopy in the crossover (see \ref{ssec_gap_crossover}), is
related to the fact that for \Li both $a$ and $a'$ are very large.
In this case all resonant interactions are unitarity-limited, so
that differential interaction shifts are absent.

Regarding the sensitivity of RF spectroscopy to small interaction
effects, which typically occur on the kHz scale or even below, \Li
features an important practical advantage over \K. In the relevant
magnetic-field region around 834\,G the \Li RF-transition
frequency changes by $-5.6$\,kHz/G, in contrast to 170\,kHz/G near
the 202\,G Feshbach resonance used in the crossover experiments in
\K. This large difference results from decoupling of the nuclear
spin from the electron spin in \Li at high magnetic fields. The
fact that a strongly interacting Fermi gas of \Li is about 30
times less susceptible to magnetic field imperfections, like
fluctuations, drifts, and inhomogeneities, facilitates precise
measurements of small interaction effects.

\subsection{Rf spectroscopy on weakly bound molecules}
\label{ssec_gap_mols}

The application of RF spectroscopy to measure binding energies of
ultracold molecules was introduced by the JILA group in
Ref.~\cite{Regal2003cum}. We have applied RF spectroscopy to
precisely determine the molecular energy structure of \Li, which
also yields precise knowledge of the two-body scattering
properties \cite{Bartenstein2005pdo}. Meanwhile, RF spectroscopy
has found various applications to ultracold Feshbach molecules
\cite{Moritz2005cim,Stoferle2006mof,Thalhammer2006llf,Ospelkaus2006uhm}.

\begin{figure}
\vspace{3mm}
\begin{center}
\includegraphics{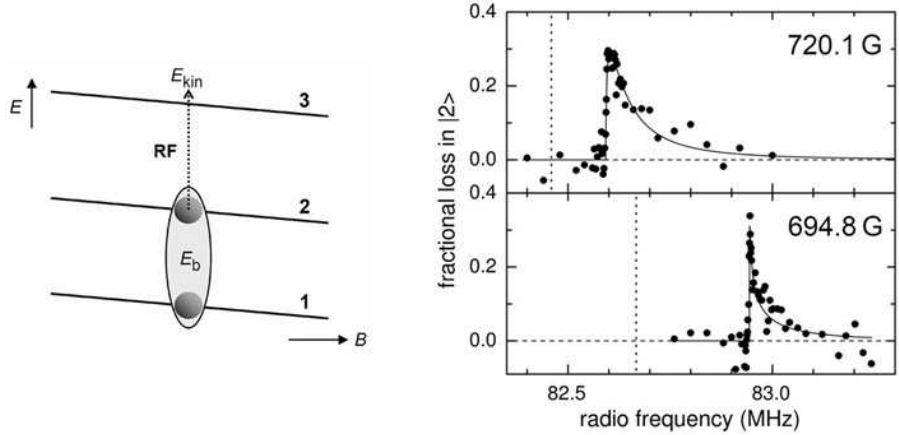}
\end{center}
\vspace{-5mm} \caption{Radio-frequency spectroscopy in the
molecular regime: basic principle (left-hand side) and
experimental results  for \Li \cite{Bartenstein2005pdo}
(right-hand side). To dissociate a molecule, the RF photon with
energy $h\nu$ has to provide at least the molecular binding energy
\Eb in addition to the bare transition energy $h\nu_0$.
In the experimental spectra, the onset of dissociation thus occurs
shifted from the bare atomic transition frequencies, which for the
two different magnetic fields are indicated by the dashed vertical
lines. The solid curves show fits by the theoretical dissociation
lineshapes according to Eq.~\ref{eq_rflineshape}.}
\label{boundfree}
\end{figure}

The basic idea of RF spectroscopy applied to weakly bound
molecules is illustrated on the left-hand side of
Fig.~\ref{boundfree}. Transferring an atom from state 2 to state 3
breaks up the dimer. The RF photon with energy $h\nu_{\rm RF}$ has
to provide at least the molecular binding energy \Eb in addition
to the bare transition energy $h\nu_{23}$. Therefore, the
dissociation sets in sharply at a threshold $\nu_{23} + E_{\rm
b}/h$. Above this threshold, the RF couples molecules to atom
pairs in the continuum with a kinetic energy $E_{\rm kin} = E -
E_{\rm b}$, where $E = h(\nu_{\rm RF} - \nu_{23})$.

The dissociation lineshape can be understood in terms of the
wavefunction overlap of the molecular state with the continuum.
For weakly bound dimers, where \Eb$=\hbar/(ma^2)$
(Eq.~\ref{eq_eb}), this lineshape is described by
\cite{Chin2005rft}
\begin{equation}
 f(E) \propto E^{-2} (E-E_{\rm b})^{1/2} (E-E_{\rm b}+E')^{-1} \, ,
\label{eq_rflineshape}
\end{equation}
where $E'=\hbar/(ma'^2)$ is an energy associated with the
(positive or negative) scattering length $a'$ between states 1 and
3. The energy $E'$ becomes important when $a'$ is comparable to
$a$, i.e.\ when both scattering channels show resonant behavior.
This is the case for \Li \cite{Gupta2003rfo,Bartenstein2005pdo},
but not for \K \cite{Regal2003cum}. In Fig.~\ref{boundfree}
(right-hand side) we show two RF-dissociation spectra taken at
different magnetic fields \cite{Bartenstein2005pdo}. The spectra
show both the change of the binding energy \Eb and the variation
of the lineshape (parameter $E'$) with the magnetic field. The
experimental data is well fit by the theoretical lineshapes of
Eq.~\ref{eq_rflineshape}.

To precisely determine the scattering properties of \Li
\cite{Bartenstein2005pdo}, we used measurements of the binding
energy \Eb in the (1,2) channel obtained through RF-induced
dissociation spectroscopy described above. In addition, we also
identified bound-bound molecular transitions at magnetic fields
where the channel (1,3) also supports a weakly bound molecular
level ($a'>0)$. These transitions do not involve continuum states
and are thus much narrower than the broad dissociation spectra.
This fact facilitated very precise measurements of magnetic field
dependent molecular transition frequencies. The combined
spectroscopic data from bound-free and bound-bound transitions
provided the necessary input to adjust the calculations based on a
multi-channel quantum scattering model by our collaborators at
NIST. This led to a precise characterization of the two-body
scattering properties of \Li in all combinations of the loweset
three spin states. This included the broad Feshbach resonance in
the (1,2) channel  at 834\,G (see discussion in
\ref{ssec_tunability}) and further broad resonances in the
channels (1,3) and (2,3) at 690\,G and 811\,G, respectively.

\subsection{Observation of the pairing gap in the crossover}
\label{ssec_gap_crossover}

After having discussed the application of RF spectroscopy to
ultracold molecules in the preceding section, we now turn our
attention to pairing in the many-body regime of the BEC-BCS
crossover. The basic idea remains the same: Breaking pairs costs
energy, which leads to corresponding shifts in the RF spectra. We
now discuss our results of Ref.~\cite{Chin2004oop}, where we have
observed the ``pairing gap'' in a strongly interacting Fermi gas.
Spectral signatures of pairing have been theoretically considered
in
Refs.~\cite{Torma2000lpo,Kinnunen2004sos,Buchler2004sos,Kinnunen2004pga,He2005rfs,Ohashi2005spe}.
A clear signature of the pairing process is the emergence of a
{\em double-peak structure} in the spectral response as a result
of the coexistence of unpaired and paired atoms. The pair-related
peak is located at a higher frequency than the unpaired-atoms
signal.

The important experimental parameters are temperature, Fermi
energy, and interaction strength. The temperature $T$ can be
controlled by variation of the final laser power of the
evaporation ramp. Lacking a reliable method to determine the
temperature $T$ of a deeply degenerate, strongly interacting Fermi
gas in a direct way, we measured the temperature $T'$ after an
isentropic conversion into the BEC limit. Note that, for a deeply
degenerate Fermi gas, the true temperature $T$ is substantially
below our temperature parameter $T'$
\cite{Carr2004aab,Chen2005toi}. The Fermi energy $E_F$ can be
controlled after the cooling process by an adiabatic recompression
of the gas. The interaction strength is varied, as in our
experiments described before, by slowly changing the magnetic
field to the desired final value.

\begin{figure}
\vspace{3mm}
\begin{center}
\includegraphics[width=12cm]{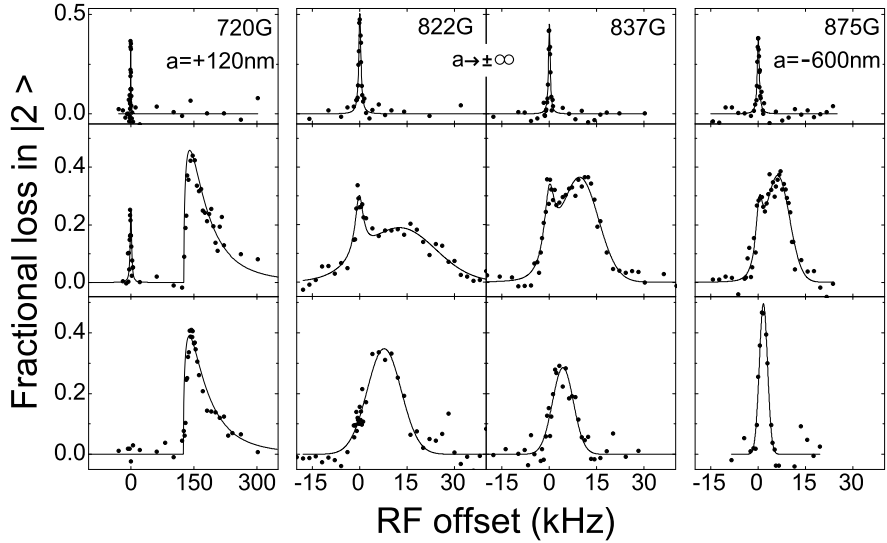}
\end{center}
\vspace{-2mm} \caption{RF spectra for various magnetic fields and
different degrees of evaporative cooling. The RF offset ($k_{\rm
B}\times 1\,\mu{\rm K} \simeq h\times20.8\,$kHz) is given relative
to the atomic transition $2\rightarrow 3$. The molecular limit is
realized for $B=720$\,G (first column). The resonance regime is
studied for $B=822$\,G and 837\,G (second and third column). The
data at 875\,G (fourth column) explore the crossover on the BCS
side. Upper row, signals of unpaired atoms at $T'\approx 6T_{\rm
F}$ ($T_{\rm F} =15\,\mu$K); middle row, signals for a mixture of
unpaired and paired atoms at $T' = 0.5T_{\rm F}$ ($T_{\rm F}
=3.4\,\mu$K); lower row, signals for paired atoms at $T' <
0.2T_{\rm F}$ ($T_{\rm F} =1.2\,\mu$K). Note that the true
temperature $T$ of the atomic Fermi gas is below the temperature
$T'$ which we measure in the BEC limit (see text). The solid lines
are introduced to guide the eye.} \label{rfmanyspectra}
\end{figure}

We recorded the RF spectra shown in Fig.~\ref{rfmanyspectra} for
different temperatures and in various coupling regimes. We studied
the molecular regime at $B=720$\,G ($a = +2170\,a_0$). For the
resonance region, we examined two different magnetic fields 822\,G
($+33,000\,a_0$) and 837\,G ($-150,000\,a_0$), because the exact
resonance location ($834.1\pm1.5$\,G, see \ref{ssec_tunability})
was not exactly known at the time of our pairing gap experiments.
We also studied the regime beyond the resonance with large
negative scattering length at $B=875$\,G ($a\approx-12,000\,a_0$).
Spectra taken in a ``hot'' thermal sample at $T\approx6T_{\rm F}$
($T_{\rm F} = 15\mu$K) show the narrow atomic $2 \rightarrow 3$
transition line (upper row in Fig.~\ref{rfmanyspectra}) and serve
as a frequency reference. We present our spectra as a function of
the RF offset with respect to the bare atomic transition
frequency.

To understand the spectra both the homogeneous lineshape of the
pair signal \cite{Kinnunen2004sos} and the inhomogeneous line
broadening due to the density distribution in the harmonic trap
need to be taken into account \cite{Kinnunen2004pga}. As an effect
of inhomogeneity, fermionic pairing due to many-body effects takes
place predominantly in the central high-density region of the
trap, and unpaired atoms mostly populate the outer region of the
trap where the density is low
\cite{Perali2004bbc,Kinnunen2004pga,Stajic2005dpo}. The spectral
component corresponding to the pairs shows a large inhomogeneous
broadening in addition to the homogeneous width of the
pair-breaking signal. For the unpaired atoms the homogeneous line
is narrow and the effects of inhomogeneity and mean-field shifts
are negligible. These arguments explain why the RF spectra in
general show a relatively sharp peak for the unpaired atoms
together with a broader peak attributed to paired atoms.

We observed a clear double-peak structure at $T'/T_{\rm F} = 0.5$
(middle row in Fig.~\ref{rfmanyspectra}, $T_{\rm F} =3.4\mu$K). In
the molecular regime (720\,G), the sharp atomic peak was well
separated from the broad dissociation signal; see discussion in
\ref{ssec_gap_mols}. As the scattering length was tuned to
resonance, the peaks began to overlap. In the resonance region
(822 and 837\,G), we still observed a relatively narrow atomic
peak at the original position together with a pair signal. For
magnetic fields beyond the resonance, we could resolve the
double-peak structure for fields up to $\sim$900\,G.

For $T'/T_{\rm F} < 0.2$, we observed a disappearance of the
narrow atomic peak in the RF spectra (lower row in
Fig.~\ref{rfmanyspectra}, $T_{\rm F} = 1.2\mu$K). This showed that
essentially all atoms were paired. In the BEC regime ($720\,$G)
the dissociation lineshape is identical to the one observed in the
trap at higher temperature and Fermi energy. Here the localized
pairs are molecules with a size much smaller than the mean
interparticle spacing, and the dissociation signal is independent
of the density. In the resonance region (822 and 837\,G) the
pairing signal showed a clear dependence on the density, which
became even more pronounced beyond the resonance (875\,G).

\begin{figure}
\vspace{3mm}
\begin{center}
\includegraphics{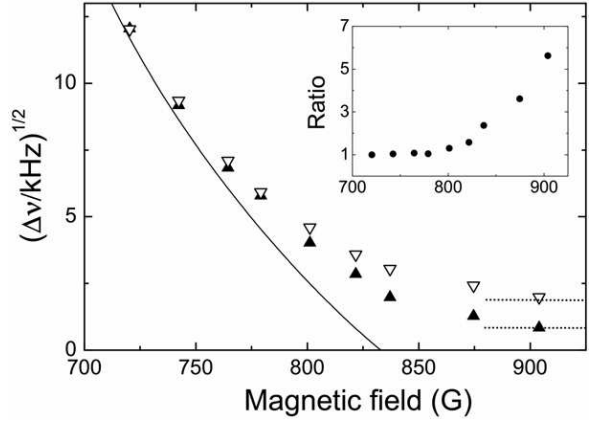}
\end{center}
\vspace{-2mm} \caption{Measurements of the effective pairing gap
$\Delta\nu$ as a function of the magnetic field $B$ for deep
evaporative cooling and two different Fermi temperatures $T_{\rm
F}=1.2\mu$K (filled symbols) and $3.6\mu$K (open symbols). The
solid line shows $\Delta\nu$ for the low-density limit, where it
is essentially given by the molecular binding energy
\cite{Eb_gap}. The two dotted lines at higher magnetic fields
correspond to the condition $2\Delta\nu = \omega_{\rm c}$ for the
coupling of the compression mode (\ref{ssec_breakdown}) to the gap
at our two different trap settings. The inset displays the ratio
of the effective pairing gaps measured at the two different Fermi
energies.} \label{gapincrossover}
\end{figure}

To quantitatively investigate the crossover from the two-body
molecular regime to the fermionic many-body regime we measured the
pairing energy in a range between 720\,G and 905\,G. The
experiments were performed after deep evaporative cooling
($T'/T_{\rm F} < 0.2$) for two different Fermi temperatures
$T_{\rm F}=1.2\,\mu$K and $3.6\,\mu$K (Fig.~\ref{gapincrossover}).
As an {\em effective pairing gap} we define $\Delta\nu$ as the
frequency difference between the pair-signal maximum and the bare
atomic resonance. In the BEC limit, the effective pairing gap
$\Delta\nu$ simply reflects the molecular binding energy $E_{\rm
b}$, as shown by the solid line in Fig.~\ref{gapincrossover}
\footnote{The maximum of the dissociation signal, which defines $h
\Delta \nu$ in the molecular regime, varies between $E_{\rm B}$
and $(4/3)\,E_{\rm B}$, depending on $E'/E_{\rm b}$ in
Eq.~\ref{eq_rflineshape}. The solid line takes this small
variation into account \cite{Eb_gap}.}. With increasing magnetic
field, in the BEC-BCS crossover, $\Delta\nu$ showed an increasing
deviation from this low-density molecular limit and smoothly
evolved into a density-dependent many-body regime where
$h\Delta\nu < E_{\rm F}$.

A comparison of the pairing energies at the two different Fermi
energies (inset in Fig.~\ref{gapincrossover}) provides further
insight into the nature of the pairs. In the BEC limit,
$\Delta\nu$ is solely determined by $E_{\rm b}$ and thus does not
depend on $E_{\rm F}$. In the universal regime on resonance,
$E_{\rm F}$ is the only energy scale and we indeed observed the
effective pairing gap $\Delta\nu$ to increase linearly with the
Fermi energy (see Ref.~\cite{BartensteinPhD} for more details). We
found a corresponding relation $h\Delta\nu \approx 0.2\,E_{\rm F}$
\footnote{Note that there is a quantitative deviation between this
experimental result for the unitarity limit (see also
\cite{BartensteinPhD}) and theoretical spectra
\cite{Kinnunen2004pga,He2005rfs,Ohashi2005spe}, which suggest
$h\Delta\nu \approx 0.35\,E_{\rm F}$. This discrepancy is still an
open question. We speculate that interactions between atoms in
state 1 and 3 may be responsible for this, which have not been
fully accounted for in theory.}. Beyond the resonance, where the
system is expected to change from a resonant to a BCS-type
behavior, $\Delta\nu$ is found to depend more strongly on the
Fermi energy and the observed gap ratio further increases. We
interpret this in terms of the increasing BCS character of
pairing, for which an exponential dependence $h\Delta\nu/E_{\rm F}
\propto \exp(-\pi/2k_{\rm F}|a|)$ (see Table \ref{tab_crossover})
is expected.

\begin{figure}
\vspace{3mm}
\begin{center}
\includegraphics{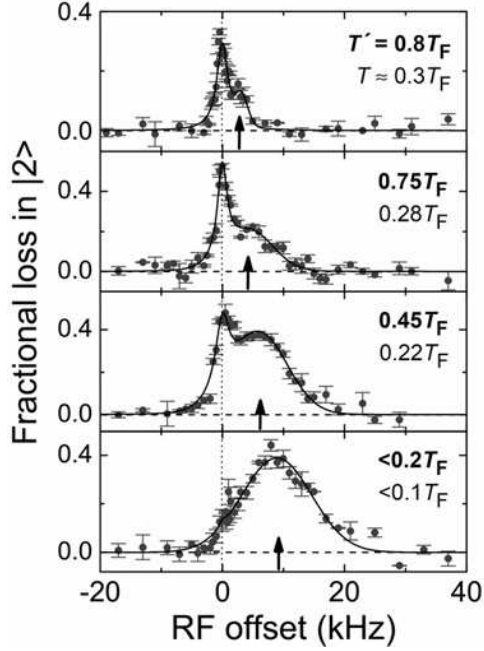}
\end{center}
\vspace{-2mm} \caption{RF spectra measured at $B=837$\,G, i.e.\
very close to the unitarity limit, for different temperatures
($T_{\rm F} = 2.5\,\mu$K). The temperature parameter $T'$ was
determined by measurements in the mBEC regime after an isentropic
conversion of the gas. Based on the entropy calculations of
Ref.~\cite{Chen2005toi} we also provide estimates for the true
temperature $T$. The solid lines are fits to guide the eye using a
Lorentzian curve for the atom peak and a Gaussian curve for the
pair signal. The vertical dotted line marks the atomic transition
and the arrows indicate the effective pairing gap $\Delta\nu$.}
\label{gapvstemp}
\end{figure}

In another series of measurements (Fig.~\ref{gapvstemp}), we
applied a controlled heating method to study the temperature
dependence of the gap in a way which allowed us to keep all other
parameters constant. After production of a pure molecular BEC
($T'< 0.2 T_{\rm F}$) in the usual way, we adiabatically changed
the conditions to $B=837$\,G and $T_{\rm F} = 1.2\,\mu$K. We then
increased the trap laser power by a factor of nine ($T_{\rm F}$
increased to $2.5\mu$K) using exponential ramps of different
durations. For fast ramps this recompression is non-adiabatic and
increases the entropy. By variation of the ramp time, we explore a
range from our lowest temperatures up to $T'/T_{\rm F} = 0.8$. The
emergence of the gap with decreasing temperature is clearly
visible in the RF spectra (Fig.~\ref{gapvstemp}). The marked
increase of $\Delta\nu$ for decreasing temperature is in good
agreement with theoretical expectations for the pairing gap energy
\cite{Chen2005bbc}.

Our pairing gap experiments were theoretically analyzed in
Refs.~\cite{Kinnunen2004pga,He2005rfs,Ohashi2005spe}. The
calculated RF spectra are in agreement with our experimental
results and demonstrate how a double-peak structure emerges as the
gas is cooled below $T/T_{\rm F}\approx0.5$ and how the atomic
peak disappears with further decreasing temperature. In
particular, the theoretical work clarifies the role of the
pseudo-gap regime \cite{Chen2005bbc,Stajic2004nos} in our
experiments, where pairs are formed before superfluidity is
reached. We believe that, the upper spectrum of
Fig.~\ref{gapvstemp} ($T' = 0.8\,$\TF, corresponding to $T =
0.3\,$\TF \cite{Chen2005toi}) shows the pseudo-gap regime. The
lower spectrum, however, which was taken at a much lower
temperature ($T' < 0.2\,$\TF, $T < 0.1\,$\TF), is deep in the
superfluid regime. Here, the nearly complete disappearance of the
atom peak shows that fermionic pairing took place even in the
outer region of the trapped gas where the density and the local
Fermi energy are low. According to theory
\cite{Kinnunen2004pga,He2005rfs} this happens well below the
critical temperature for the formation of a resonance superfluid
in the center of the trap. This conclusion \cite{Chin2004oop} fits
well to the other early observations that suggested superfluidity
 in experiments performed under similar conditions \cite{Kinast2004efs,Bartenstein2004ceo},
and also to the observation of superfluidity by vortex formation
in Ref.~\cite{Zwierlein2005vas}.

We finally point to an interesting connection to our measurements
on radial collective excitations (\ref{ssec_breakdown}), where an
abrupt breakdown of hydrodynamics was observed at a magnetic field
of about 910\,G \cite{Bartenstein2004ceo}. The hydrodynamic
equations which describe collective excitations implicitly assume
a large gap, and their application becomes questionable when the
gap is comparable to the radial oscillation frequency
\cite{Combescot2004coc}. We suggest a pair breaking condition
$\omega_{\rm c} = 2h\Delta\nu$~\footnote{The factor of $2$ in this
condition results from the fact that here pair breaking creates
two in-gap excitations, instead of one in-gap excitation in the
case of RF spectroscopy, where one particle is removed by transfer
into an empty state.}, which  roughly corresponds to $\omega_{\rm
r} = h\Delta\nu$ ($\omega_{\rm c} \approx 2\omega_{\rm r}$).  Our
pair breaking condition is illustrated by the dashed lines in
Fig.~\ref{gapincrossover} for the two different Fermi energies of
the experiment. In both cases the effective gap $\Delta\nu$
reaches the pair breaking condition somewhere slightly above
900\,G. This is in striking agreement with our observations on
collective excitations at various Fermi temperatures
\cite{Bartenstein2004ceo,BartensteinPhD}. This supports the
explanation that pair breaking through coupling of oscillations to
the gap leads to strong heating and large damping and thus to a
breakdown of superfluidity on the BCS side of the resonance.

\section{Conclusion and outlook}
\label{sec_outlook}

Ultracold Fermi gases represent one of the most exciting fields in
present-day physics.
Here experimental methods of atomic, molecular, and optical
physics offer unprecedented possibilities to explore fundamental
questions related to many different fields of physics.

In the last few years, we have seen dramatic and also surprising
developments, which have already substantially improved our
understanding of the interaction properties of fermions. Amazing
progress has been achieved in the exploration of the crossover of
strongly interacting system from BEC-type to BCS-type behavior.
Resonance superfluidity now is well established. Recent
experimental achievements have made detailed precision tests of
advanced many-body quantum theories possible.

The majority of experiments have so far been focused on bulk
systems of two-compo\-nent spin mixtures in macroscopic traps;
however, ultracold gases offer many more possibilities to realized
intriguing new situations. The recent experiments on imbalanced
systems \cite{Partridge2005pap,Zwierlein2005fsw} give us a first
impression how rich the physics of fermionic systems will be when
more degrees of freedom will be present. As another important
example, optical lattices
\cite{Modugno2003poa,Kohl2005fai,Chin2006efs} allows us to model
the periodic environment of crystalline materials, providing
experimental access to many interesting questions in
condensed-matter physics \cite{Lewenstein2006uag}. Also, fermionic
mixtures of different atomic species open up many new
possibilities. In Bose-Fermi mixtures
\cite{Modugno2002coa,Stan2004oof,Inouye2004ooh,Ospelkaus2006uhm},
fermionic pairing and superfluidity can be mediated through a
bosonic background \cite{Bijlsma2000pei,Efremov2002pwc}. In the
case of Fermi-Fermi mixtures \cite{Kerner2006bh}, pairing between
particles of different masses \cite{Petrov2006dmi} and novel
regimes of superfluidity represent intriguing prospects for future
research.

With the recent experiments on the physics of ultracold fermions,
we have just opened the door to an exciting new research field. On
the large and widely unexplored terrain, many new challenges (and
surprises) are surely waiting for us!


\acknowledgments

Our work on ultracold Fermi gases, which developed in such an
 exciting way, is the result of a tremendous team effort
over the past eight years. Its origin dates back to my former life
in Heidelberg (Germany), where our activities on ultracold
fermions began in the late 1990´s. I thank the team of these early
days (A.~Mosk, M.~Weidem\"uller, H.~Moritz, T.~Els\"asser) for the
pioneering work to start our adventures with \Li. The experiment
moved to Innsbruck in 2001, and many people have contributed to
its success there. For their great work and achievements, I thank
the Ph.D.\ students S.~Jochim (who moved with the experiment from
Heidelberg to Innsbruck) and M.~Bartenstein, A.~Altmeyer, and
S.~Riedl, along with the diploma students G.~Hendl and
C.~Kohstall. Also, I acknowledge the important contributions by
the post-docs R.~Geursen and M.~Wright (thanks, Matt, also for the
many useful comments on the manuscript). I am greatly indebted to
C.~Chin, who shared a very exciting time with us and stimulated
the experiment with many great ideas, and my long-standing
colleague J.~Hecker~Denschlag for their invaluable contributions.
The experiment strongly benefited from the great synergy in a
larger group (\verb"www.ultracold.at") and from the outstanding
scientific environment in Innsbruck. Finally, I thank the Austrian
Science Fund FWF for funding the experiment through various
programs, and the European Union for support within the Research
Training Network ``Cold Molecules''.


\bibliographystyle{varenna4}
\bibliography{ultracold,notes}

\end{document}